\documentclass[twocolumn,pra,superscriptaddress]{revtex4-1}

\usepackage[pdftex,
			pdftitle={Linear-and-quadratic reservoir engineering of non-Gaussian states},
			pdfauthor={Oussama Houhou},
			pdfpagemode=UseOutlines,
			pdfstartview=FitH]{hyperref}
\usepackage{amsmath}
\usepackage{amsfonts}
\usepackage{natbib}
\usepackage[utf8]{inputenc}
\usepackage[sans]{dsfont}
\usepackage[dvipsnames]{color}
\usepackage{marvosym}
\usepackage{graphicx}
\usepackage{enumerate}

\usepackage{bbm}

\newcommand{\ket}[1]{\ensuremath{\left|#1\right\rangle}}
\newcommand{\bra}[1]{\ensuremath{\left\langle #1\right|}}
\newcommand{\braket}[2]{\ensuremath{\left\langle #1\left|\right.#2\right\rangle}}

\newcommand{\diff}{\ensuremath{\mathrm{d}}}

\newcommand{\EXP}[1]{\ensuremath{e^{#1}}}

\newcommand{\hc}[1]{\ensuremath{\mathrm{H.c.}}}

\newcommand{\dB}{\ensuremath{~\mathrm{dB}}}

\newcommand{\refeq}[1]{Eq.~(\ref{#1})}
\newcommand{\Refeq}[1]{(\ref{#1})}
\newcommand{\reffig}[1]{Fig.~\ref{#1}}

\newcommand{\refsec}[1]{Section~\ref{#1}}
\newcommand{\refappndx}[1]{Appendix~\ref{#1}}

\newcommand{\Ai}{{\mbox{Ai}}}

\newcommand{\laguerre}{\ensuremath{\mathrm{L}}}
\newcommand{\Hermite}{\ensuremath{\mathrm{H}}}

\newcommand{\displ}{\ensuremath{\hat D}}
\newcommand{\sqz}{\ensuremath{\hat S}}

\renewcommand{\refsec}[1]{Sec.~\ref{#1}}

\begin{document}

\title{Linear-and-quadratic reservoir engineering of non-Gaussian states}
\author{Matteo Brunelli}
\affiliation{Cavendish Laboratory, University of Cambridge, Cambridge CB3 0HE, United Kingdom}

\author{Oussama Houhou}
\affiliation{Centre for Theoretical Atomic, Molecular, and Optical Physics, School of Mathematics and Physics, Queen's University, Belfast BT7 1NN, United Kingdom}
\affiliation{Laboratory of Physics of Experimental Techniques and Applications, University of Medea, Medea 26000, Algeria}

\begin{abstract}
	We study the dissipative preparation of pure non-Gaussian states of a target mode which is coupled both linearly and quadratically to an auxiliary damped mode. We show that any pure state achieved independently of the initial condition is either (i) a cubic phase state, namely a state given by the action of a non-Gaussian (cubic) unitary on a squeezed vacuum or (ii) a (squeezed and displaced) finite superposition of Fock states. Which of the two states is realized depends on whether the transformation induced by the engineered reservoir on the target mode is canonical (i) or not (ii). We discuss how to prepare  these states in an optomechanical cavity driven with multiple control lasers, by tuning  the relative strengths and phases of the drives. Relevant examples in (ii) include the stabilization of 
	 mechanical  Schr\"odinger cat-like states or Fock-like states of any order. 
Our analysis is entirely analytical, it extends reservoir engineering to the non-Gaussian regime and enables the preparation of novel mechanical states with negative Wigner function.
\end{abstract}

\maketitle


\section*{Introduction}
In recent groundbreaking experiments, the autonomous stabilization of single- and two-mode mechanical squeezed states  
has been achieved
\cite{wollman2015quantum,pirkkalainen2015squeezing,lecocq2015quantum,ockeloen2018stabilized}.
For a single mode, mechanical squeezing is simply obtained by driving an optomechanical cavity with two control lasers with unequal amplitudes~\cite{clerk2013squeezing}.
	The bichromatic drive effectively couples the cavity mode to a Bogoliubov  mode of the target resonator, so that  cavity cooling of the Bogoliubov mode results in the desired squeezing. Crucially, this simple scheme does not rely on measurement-and-feedback loops; the target system rather relaxes into a squeezed steady state irrespectively of its initial state. This scheme pertains to a set of techniques, commonly referred to as reservoir engineering, to stabilize genuine quantum features of a system by tailoring the properties of the environment~\cite{poyatos1996quantum,wang2013reservoir,woolley2014two}. In this respect, a damped cavity mode provides a highly tunable reservoir, where different system-environment couplings can be engineered by a suitable choice of the drives.  Besides cavity optomechanics, reservoir engineering has been successfully applied to trapped atoms~\cite{krauter2011entanglement}, ions~\cite{barreiro2011open,lin2013dissipative,kienzler2015quantum}, and to circuit quantum electrodynamics~\cite{shankar2013autonomously,leghtas2015confining}.

	A major advance for bosonic reservoir engineering would be the stabilization of non-Gaussian states, which requires the implementation of  a nonlinear transformation of the target mode. As first suggested in Refs.~\cite{poyatos1996quantum,de1996even} for trapped ions, a coupling which is {\it quadratic} in the target mode can be exploited for the dissipative preparation of a Schr\"odinger cat state. However, unlike for \mbox{reservoir-engineered} squeezing, the steady state in this case is  no longer independent of the initial condition~\cite{tan2013generation,asjad2014reservoir}. The protocol necessitates initialization of the target system into often prohibitive states (e.g. a single Fock state), which undermines the very purpose of reservoir engineering.

	\begin{figure*}[t!]
		\centering
		\includegraphics[scale=.65]{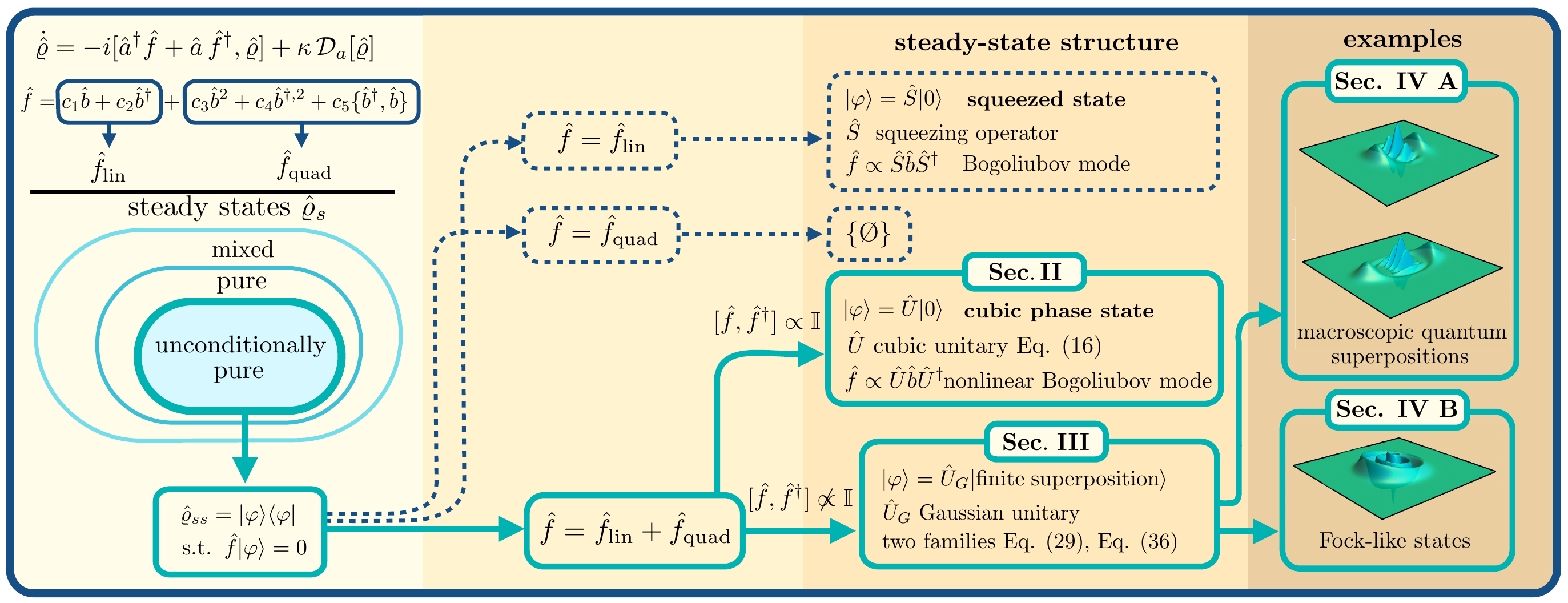}
		\caption{Synoptic scheme of the paper. As shown on the left panel, we study the open dynamics of two interacting bosonic modes -- the target ($\hat b$) and the auxiliary ($\hat a$) mode. The coupling has both a linear and a quadratic part in $\hat b$ and the coefficients $c_j,\;j=1,\ldots,5$ can be independently tuned. Depending on the choice of the initial state and the coefficients, either pure or mixed steady states can be stabilized. Among them, we focus on {\it unconditionally pure states} of mode $\hat b$, namely pure states that are reached asymptotically regardless of the initial condition. These states must be annihilated by the operator $\hat f=\hat f(b,b^\dagger)$ and there must be no conserved quantities. For the case of a linear transformation $\hat f=\hat f_{\mathrm{lin}}$, unconditionally pure states coincide with (rotated) squeezed states, while there are no such states for a purely quadratic  transformation  $\hat f=\hat f_{\mathrm{quad}}$; these cases are displayed in the dashed boxes. On the other hand, for the transformation $\hat f=\hat f_{\mathrm{lin}}+\hat f_{\mathrm{quad}}$ any unconditionally pure state belongs to either one of two distinct classes, depending on the choice of
		$c_j$. If the coefficients are such that $\hat f$ {\it is bosonic}, then the steady state is a {\it cubic phase state}, namely the state obtained by the action of a non-Gaussian (cubic) unitary on a squeezed vacuum; otherwise it is written as a Gaussian unitary acting on a {\it finite superposition} of Fock states. All these states are non-Gaussian, and in particular non-classical. Furthermore, two distinct families of unconditionally pure states can be found in (ii). Relevant examples are macroscopic quantum superposition states and states that can approximate with arbitrary precision any (displaced) Fock state.
		\label{f:Overview}}
	\end{figure*}

	In this work we show how pure non-Gaussian states of a target system can be unconditionally prepared, i.e., {\it without requiring  initialization}, by engineering a coupling with the environment (cavity field) that is {\it both linear and quadratic} in the target mode. Specifically, the target states we consider are right eigenstates (with zero eigenvalue) of the combination $\hat f(\hat b,\hat b^\dagger)=\hat f_{\mathrm{lin}}+\hat f_{\mathrm{quad}}$, where $\hat f_{\mathrm{lin}}\,(\hat f_{\mathrm{quad}})$ represents a generic linear (quadratic) function of the target mode $\hat b$. These states meet all the {\it desiderata}, inasmuch they are pure, non-Gaussian and  attained unconditionally. In particular, being pure non-Gaussian states, {\it they all possess a non-positive Wigner representation}~\cite{hudson1974wigner}. Linear-and-quadratic reservoir engineering thus provides a distinct advantage over the purely quadratic case $\hat f(\hat b,\hat b^\dagger)=\hat f_{\mathrm{quad}}$ for the same degree of nonlinearity. Loosely speaking, the quadratic part of the transformation provides the non-Gaussian resource, while the insensitiveness of the steady state to the initial condition is restored by the linear part, which breaks the symmetry under parity transformation. 

	Furthermore, we prove that any state possessing these features can be of only two kinds: (i)~a cubic phase state, namely a state obtained by the action of a cubic gate on a squeezed vacuum or (ii)~a Gaussian unitary (squeezing and displacement) acting on a  superposition of a {\it finite number of Fock states}. State (i) plays a fundamental role in continuous variable quantum computation, where it allows to realize any arbitrary unitary operation and renders measurement-based quantum computation universal~\cite{gottesman2001encoding,ghose2007non}; states (ii) represent a novel family of bosonic states. These two classes single out distinct ways of allocating the non-Gaussian resource---either in the unitary operation or in the finiteness of the superposition---which are determined by whether the transformed operator $\hat f$ is bosonic (i) or not (ii). Our analysis is fully analytical and provides exact expressions for all these states.

	We then show how the required linear-and-quadratic coupling can be implemented in a cavity that is parametrically coupled to both the mechanical displacement and the displacement squared. Driving the cavity with multiple control drives and tuning their relative strengths and phases, the mechanical resonator can be stabilized  in the states (i) and (ii). In particular, we show that within class (ii) it is possible to stabilize novel macroscopic superpositions similar to a Schr\"odinger cat state and 
	to approximate \emph{any} (displaced) Fock state with arbitrary precision.
	Our scheme extends reservoir engineering of squeezing to the non-Gaussian regime and enables the unconditional preparation of states with negative Wigner function. The present work substantially extends the previous findings of Ref.~\cite{brunelli2018unconditional} in the following sense: it addresses in an exhaustive way the pure states that can be stabilized by linear-and-quadratic resources, i.e., the solutions provided are all and the only admissible. One of these,
	namely Eq.~\eqref{SecondCatLike}, was already discussed by us in an optomechanical setting \cite{brunelli2018unconditional}. For the sake of generality, we first develop our analysis at an abstract level, and then discuss in detail a proposed optomechanical realization.
	
	This work is structured as follows. In Sec.~\ref{s:Model} we introduce the model, describe the steady-state structure and revise some known cases. The first kind of target state, the cubic phase state, is discussed in Sec.~\ref{s:CubicPhase}, while the second kind, consisting of two families of finite superpositions, in Sec.~\ref{s:FiniteSup}; relevant examples within the second class are given in Sec.~\ref{s:ExFiniteSup}. Sec.~\ref{s:Imprecision} contains an analysis of the major sources of imperfection affecting the target states, namely the finite accuracy in tuning the value of the coefficients and both quantum and classical noise acting on the target system.  In Sec.~\ref{s:Optomech} we discuss how to implement our protocol in an optomechanical setup. Finally, in Sec.~\ref{s:Conclusions} we draw the conclusions of our work. 
	An overview of the core of the paper is provided in Fig.~\ref{f:Overview} to help the navigation through Sec.~\ref{s:Model} to~\ref{s:ExFiniteSup}---which constitute the more technical  part of our work---and highlight the main results.


\section{The model and steady-state structure}\label{s:Model}
	We consider two interacting bosonic modes---a target  and an auxiliary mode---that we label $\hat b$ and $\hat a$, respectively. The interaction between them takes the form
	\begin{equation}\label{BS}
		\hat H=\hat a^\dagger \hat f +\hat a \hat f^\dagger \, ,
	\end{equation}
	where $\hat f=\hat f(\hat b,\hat b^{\dag})$ is an at most quadratic  (but otherwise general)  function of the target mode $\hat b$, i.e.,
	\begin{equation}\label{fmode}
		\hat f=c_1 \hat b+c_2 \hat b^{\dag} +c_3 \hat b^2+c_4 \hat b^{{\dag}\,2}+c_5\{\hat b^{\dag} ,\hat b \} \, ,
	\end{equation} 
	whose complex coefficients $c_j,\;j=1,\ldots,5$ can be independently tuned. Here and in the rest of the work $\{ \cdot ,\cdot \}$ denotes the anti-commutator and we set $\hbar=1$. Equation \Refeq{BS} describes a beam-splitter interaction between the auxiliary field mode and a nonlinear combination of the target creation and annihilation operators. We further assume that target system experiences negligible losses, while the auxiliary mode dissipates into an effective zero temperature reservoir. The master equation describing the evolution of the joint density matrix $\hat\varrho$ reads 
	\begin{equation}\label{MasterEq}
		\dot{\hat \varrho}=-i[\hat H,\hat \varrho]+\kappa_a\, \mathcal{D}_{a}[\hat \varrho]\equiv\mathcal{L}[\hat \varrho] \, ,
	\end{equation}
	where $\mathcal{D}_{ o}[\hat\varrho]=\hat o \hat \varrho \hat o^{\dag}-\frac12\bigl(\hat o^{\dag} \hat o\hat \varrho + \hat \varrho\hat o^{\dag} \hat o \bigr)$ is the standard dissipator with damping rate $\kappa_a$, and the unitary and dissipative terms have been grouped together in the so-called Liouvillian super-operator  $\mathcal{L}$. Similarly to cavity cooling, the interaction swaps the state of the auxiliary mode with that encoded in the nonlinear operator $\hat f$ and, as the first is in contact with a zero entropy reservoir, the second is effectively cooled. As will be shown later, for this picture to be valid, $\hat f$ itself does not need to be a bosonic mode, namely to satisfy canonical commutation relations.

	If the coefficients $c_j$ are such that a steady state exists, this fulfills the condition
	\begin{equation}\label{SteadyL}
		\mathcal{L}[\hat \varrho_{ss}]=0 \, .
	\end{equation}
	Moreover, for the evolution described in \refeq{MasterEq}, the steady state is given by $\nobreak{\hat \varrho_{ss}=\ket{\psi_{ss}}\bra{\psi_{ss}}}$, with $\nobreak{\ket{\psi_{ss}}=\ket{0}\otimes\ket{\varphi}}$ and where the target state obeys the dark state condition~\cite{kraus2008preparation}
	\begin{equation}\label{DarkState}
		\hat f \ket{\varphi}=0 \, .
	\end{equation}
	An intuitive picture to understand \refeq{DarkState} as the result of an engineered cooling process is to consider the situation where the auxiliary mode can be adiabatically eliminated (see \refsec{s:Imprecision} for more details). In this case, the steady state of the target system $\hat \varrho^{(b)}_{ss}=\mathrm{Tr}_{a}[{\hat \varrho_{ss}}]$ is defined by the condition
	\begin{equation}
		\mathcal{D}_{\hat f}\,\bigl[\hat \varrho^{(b)}_{ss}\bigr]=0\ .
	\end{equation}
	 The target system thus experiences a dissipative dynamics that cools it toward the ground state of the operator $\hat f$, in agreement with \refeq{DarkState}.
	Additional dissipation due to the presence of a finite temperature bath results in a mixed steady state for the target mode, and will be considered in Sec.~\ref{s:Imprecision}. 
	
	For convenience, we introduce  the following notation
	\begin{equation}\label{LinQuad}
		\hat f=\hat f_{\text{lin}}+\hat f_{\text{quad}} \, ,
	\end{equation}
	where $\hat f_{\text{lin}}=c_1 \hat b+c_2 \hat b^{\dag}$ and $\hat f_{\text{quad}}=c_3 \hat b^2+c_4 \hat b^{{\dag}\,2}+c_5\{\hat b^{\dag} ,\hat b \}$ collect the linear and the quadratic terms, respectively.

	Since the target state $\ket{\varphi}$ is annihilated by a coherent superposition of $\hat b,\, \hat b^\dagger$ and their powers/product, it develops coherences in the Fock basis that  are ultimately responsible for its nonclassical features. However, depending on the choice of the coefficients $c_j$, there can be none, one or two independent solutions of the dark state condition. Since $\hat \varrho_{ss}=\lim_{t\rightarrow+\infty}\EXP{\mathcal{L}t}\hat \varrho(0)$, when \refeq{DarkState} has more than one solution, different initial conditions will cause different states to be populated  in the infinite-time limit. In particular, depending on the initial state, the model described by \refeq{MasterEq} admits either pure or mixed steady states. Without initialization (e.g. starting from a thermal state), the steady state will be in general a mixture of pure non-Gaussian states, which is hardly nonclassical. On the other hand, initialization of the system to some specific state is often prohibitive. Indeed, in order to target a desired (nonclassical) steady state, the system may have to be initialized in a state that is already nonclassical, which defies the purpose of reservoir engineering~\cite{tan2013generation}. Moreover, choosing the correct initial state requires the knowledge of all the conserved quantities, a task that already in our case is not trivial~\cite{albert2014symmetries}.

	These considerations motivate us to focus on the subset of pure steady states (see Fig.~\ref{f:Overview}) that are achieved independently of the initial state. Such states correspond to single solutions of \refeq{DarkState} and in this work we will call them {\it unconditionally pure states}. Steady states insensitive to the the initial conditions are sometimes referred to as `unique' or `attractive' in the spectral theory of open systems ~\cite{schirmer2010stabilizing,nigro2018uniqueness}. However, some system may retain such property but have a steady state that changes by varying some Hamiltonian parameter, in which case the system displays multi-stability and a related dissipative phase transition (in some meaningful thermodynamic limit)~\cite{minganti2018spectral}. To avoid confusion with this meaning of (non)uniqueness, we dub unconditionally pure any pure steady state that is unique and toward which all states converge in the infinite-time limit for all values of Hamiltonian parameters. 
	
	Before we move to the characterization of the dark states of \refeq{LinQuad}, we conclude this Section by reviewing the case of either linear or quadratic reservoir engineering of a bosonic mode.


	\subsection{Linear reservoir engineering}\label{s:ResLin}
		Let us consider the case of a bilinear coupling between the auxiliary and the target system, i.e. $\hat f(\hat b,\hat b^{\dagger})=\hat f_{\mathrm{lin}}$ in \refeq{BS}. In this case \refeq{SteadyL} has always at most one solution, i.e., if a stationary state exists, it is unconditionally pure. Moreover, when $\vert c_1\vert>\vert c_2\vert$ the system admits a stable steady state. It is easy to show that this coupling always induces a canonical transformation of the target mode $\hat b$. Indeed, the interaction can be rewritten as $\hat H=\mathcal{G}(\hat a^\dagger \hat \beta +\hat a \hat\beta^\dagger )$~\cite{clerk2013squeezing}, where $\mathcal{G}=\sqrt{\vert c_1\vert^2-\vert c_2\vert^2}$ and we have introduced the  Bogoliubov mode
		\begin{equation}\label{Bogoliubov}
			\hat \beta=\mu \hat b+\nu \hat b^\dagger\, ,
		\end{equation}
		with $\mu=\cosh r$, $\nu=\EXP{i\theta}\sinh r$, $\theta\equiv \arg c_2$ and squeezing parameter defined by $\tanh r =\frac{\vert c_2\vert }{\vert c_1\vert}$; without loss of generality, we set the phase of $c_1$ to  zero, which is always possible by a suitable choice of the phase of the auxiliary mode $\hat a$. As it is well known, the Bogoliubov mode \Refeq{Bogoliubov} is obtained via the unitary action $\hat \beta=\sqz(\xi)~\hat b~\sqz^\dagger(\xi)$,  where $\sqz(\xi)=\EXP{\frac{\xi^*}{2}\hat b^2-\frac{\xi}{2}\hat b^{\dagger\,2}}$ is the single mode squeezing operator of argument $\xi=r \EXP{i \theta}$. 
		It immediately follows that $\hat \beta$ is bosonic too, i.e.,  $[\hat \beta,\hat \beta^\dagger]=\mathbbm{1}$. Also note that, due to the rescaling, the mode $\hat f$ obeys the commutation relations $[\hat f,\hat f^\dagger]=\mathcal{G}^2\mathbbm{1}$. The dark state condition then becomes $\hat b \,\sqz^\dagger \ket{\varphi}=0$, which in turn yields $\ket{\varphi}=\sqz\ket{0}$, 
		namely  the steady state of the target mode is a rotated squeezed state. The point that we want to emphasize here is that, when $\hat f$ and $\hat b$ are unitarily equivalent---and therefore $\hat f$ is bosonic---the dark state condition defining the steady state of the dissipative dynamics~\Refeq{MasterEq} can be equivalently characterized as a unitary transformation acting on the vacuum. Such an expression explicitly characterizes the steady state as an unconditionally pure state.

	\subsection{Quadratic reservoir engineering}\label{s:ResQuad}
		In case of a purely quadratic coupling $\hat f(\hat b,\hat b^{\dagger})=\hat f_{\mathrm{quad}}$, the system possesses a discrete $\mathbb{Z}_2$ symmetry, i.e., $[\hat H,\hat \Pi]=0$, where $\hat \Pi=\EXP{i\pi\hat  b^\dag \hat b}$ is the parity operator (note that the dissipator in \refeq{MasterEq} trivially commutes with $\hat \Pi$). Since the parity is conserved during the evolution, we immediately conclude that the model does not admit any unconditionally pure steady state \cite{albert2014symmetries}. This fact has a direct consequence for the dissipative preparation of Schr\"odinger cat states where, in order to stabilize an even/odd cat, the system must be initialized in a $\pm1$ eigenstate of $\hat \Pi$, e.g. $\ket{0}$, $\ket{1}$~\cite{poyatos1996quantum,tan2013generation}. For any choice of the coefficients in $\hat f_{\mathrm{quad}}$, we always have $[\hat f_{\mathrm{quad}},\hat f_{\mathrm{quad}}^{\dagger}]\not\propto  \mathbbm{1}$, so that we might be tempted to conclude that  giving up the canonical character of the transformation implies the loss of the uniqueness of the steady state. As we will show for a linear-and-quadratic transformation, this is not the case: while the action of a canonical transformation on $\hat b$ ensures the uniqueness of the steady state, the converse is not necessarily true. The condition expressed by $[\hat f,\hat f^\dagger]=\mathcal{G}^2\mathbbm{1}$ is thus only sufficient for uniqueness.

\section{The cubic phase state}\label{s:CubicPhase}
	We now move to the characterization of the unconditionally pure steady states accessible by a linear-and-quadratic coupling, i.e., the states annihilated by the nonlinear operator \refeq{fmode}. Similarly to what we have shown for the linear case, if the coefficients $c_j$ are such that $\hat b$ and $\hat{f}$ are related by a unitary transformation $\hat{f}=\mathcal{G}~\hat U\hat b \hat U^\dagger$, then the steady state $\ket{\varphi}=\hat U\ket{0}$ is unconditionally pure. We therefore enforce canonical commutation relations on the nonlinear operator $\hat f=\hat f_{\text{lin}}+\hat f_{\text{quad}}$. The condition $[\hat{f},{\hat{f}}^\dagger]=\mathcal{G}^2\mathbbm{1}$, in addition to the hyperbolic identity fulfilled  by the linear coefficients in Eq.~\Refeq{Bogoliubov}, results in the following set of equations for the coupling parameters
	\begin{eqnarray}
		|c_3|				&=&	|c_4|\ ,\label{eqn:unitary-condition-2}\\
		c_3c_5^*			&=&	c_4^*c_5\ ,\label{eqn:unitary-condition-3}\\
		c_1c_5^*+c_1^*c_3	&=&	c_2c_4^*+c_2^*c_5\ .\label{eqn:unitary-condition-4}
	\end{eqnarray}
	Using Eqs.~\Refeq{eqn:unitary-condition-2} and \Refeq{eqn:unitary-condition-3} we find $\phi_5=\frac12(\phi_3+\phi_4)+k\pi$ with $k\in\mathbb{Z}$, where we set $c_j=|c_j|\EXP{i\phi_j}$ (with $\phi_1=0$ and $\phi_2=\theta$). Moreover, \refeq{eqn:unitary-condition-4} can be put in the form
	\begin{equation}
		\frac{|c_5|}{|c_3|}=(-1)^{k+1} \,\EXP{\frac{i}{2}(3\phi_3+\phi_4)}\frac{1-\EXP{i(\theta-\phi_3-\phi_4)}\, \zeta}{1-\EXP{-i(\theta-\phi_3-\phi_4)}\, \zeta}\ ,
	\end{equation}
	 with $\zeta=\tanh r$, which is true if $|c_5|=|c_3|$ and
	\begin{equation}\label{eqn:phases-condition}
		\frac{\zeta \sin\left(-\theta+\phi_3+\phi_4\right)}{1-\zeta \cos\left(-\theta+\phi_3+\phi_4\right)}=(-1)^k\left(\tan\frac{3\phi_3+\phi_4}{2}\right)^{(-1)^{k+1}}.
	\end{equation}
	Equation \Refeq{eqn:phases-condition} should be valid for all values of $r$. Therefore, we obtain  $|c_3|=|c_4|=|c_5|$ and the two conditions
	\begin{align}
		\phi_3&=-\frac{1}{2}\left(\theta-\ell\pi\right)+\frac{\pi}{2}\left(1+(-1)^k\right)\,, \\
		\phi_4&=\frac{3}{2}\left(\theta-\ell\pi\right)-\frac{\pi}{2}\left(1+(-1)^k\right)\ ,
	\end{align}
	where $\ell$ is integer. Putting everything together, the mode $\hat{f}$ must take the form 
	\begin{equation}\label{eqn:f-parameters}
		\hat f=\mathcal{G}\left[\hat \beta-(-1)^{k +\frac{\ell}{2}} t\ \EXP{\frac{i}{2}\theta}\left(\EXP{-\frac{i}{2}\theta}\hat b-(-1)^\ell\EXP{\frac{i}{2}\theta}\hat b^\dagger\right)^2\right]\ ,
	\end{equation}
	where $\hat \beta$ is the Bogoliubov mode of \refeq{Bogoliubov} and we set $\nobreak{t\equiv|c_3|/ \mathcal{G}}$. The first term on the right-hand side of \refeq{eqn:f-parameters} corresponds to the action of a squeezing operation, while the other terms perform a cubic operation modulo a rotation. Equation \Refeq{eqn:f-parameters} defines a {\it nonlinear Bogoliubov transformation} and in the limit of vanishing nonlinearity $t\rightarrow0$ reduces to the familiar expression of \refeq{Bogoliubov}. In order for $\hat f$ to describe a (rescaled) bosonic mode, the nonlinear terms in \refeq{fmode} are forced to appear with the same magnitude and a definite relative phase to form a squared rotated quadrature; also notice that the squeezing is either parallel (odd $\ell$) or orthogonal (even $\ell$) to the nonlinear term. A less general version of quadrature-dependent Bogoliubov transformation was studied in Ref.~\cite{de2001quadrature} in the context of nonlinear quantum optics.

	As a consequence, the mode $\hat{f}$ is obtained from $\hat b$ via the action of the unitary operator
	\begin{equation}\label{UCubic}
		\hat U=\hat R(\theta/2)\hat F^{\ell+1}\hat \Gamma(\gamma)\hat F^{\dagger\,\ell+1}\hat R^\dagger(\theta/2)\sqz(\xi)\ ,
	\end{equation}
	where we have introduced a phase rotation $\hat R(\phi)=\EXP{i\phi \hat b^\dagger \hat b}$, the Fourier operator $\hat F=\hat R(\pi/2)$  and the so-called {\it cubic phase gate} $\hat \Gamma(\gamma)$  defined by $\hat \Gamma(\gamma)=\EXP{i\gamma x^3}$, where $\gamma\in\mathbb{R}$ is the {\em cubicity} parameter given by 
	\begin{equation}
		\gamma=(-1)^{k+\ell}\frac{\sqrt8 t}{3\left[\mu+(-1)^\ell\nu\right]}\ .
	\end{equation}
	The  corresponding steady state $\ket{\varphi}=\hat U\ket{0}$  is a generalized  {\it cubic phase state}~\cite{Weedbrook:12}. The cubic phase state was originally introduced in Ref.~\cite{gottesman2001encoding} as the state $\nobreak{\ket{\gamma}=\EXP{i \gamma \hat x^3 }\ket{0}_p}$, where $\ket{0}_p$ is the zero momentum eigenstate, as an off-line resource to implement the cubic phase gate $\hat \Gamma$. Such a state is however an improper eigenstate, and thus unphysical;  a correctly normalized version is actually given by the unitary operator Eq.~\eqref{UCubic} acting on the vacuum. Indeed, for $\theta=-\pi,\ \ell=0$ and any $k$ we obtain the momentum-squeezed cubic phase state
	\begin{equation}\label{eqn:cubic-phase-state-momentum}
		\ket{\gamma,r}=\hat\Gamma(\gamma)\sqz(-r)\ket{0}\,.
	\end{equation}
	It is easy to see that in the limit of infinite squeezing $\ket{\gamma}$ is recovered,  while for zero cubicity a squeezed vacuum is retrieved. 
	
	The cubic phase state is an important resource in many quantum information protocols \cite{menicucci2006universal}. Specifically, having access to this cubic non-linearity in addition to Gaussian operations (namely squeezing, rotation and displacement), it is possible to implement arbitrary unitary operations, which are key components required for quantum computation and other developing applications of quantum information processing~\cite{gu2009quantum}. For example, the cubic phase state can be used to generate a non-Gaussian cluster state, to be exploited in a complete protocol for universal measurement-based quantum computation, e.g. with mechanical oscillators in optomechanical systems \cite{houhou2018unconditional}. While other attempts so far have been focusing to the generation of approximate quantum states of light with weak cubic non-linearity \cite{marek2011deterministic,yukawa2013emulating,marshall2015repeat,miyata2016implementation,marek2018general}, our proposal is capable of preparing  unconditionally  genuine quantum states of matter or light exhibiting cubic non-linearities as big as the intrinsic system's linear and quadratic couplings allow.
	
\begin{figure}[t]
		\includegraphics[width=\columnwidth]{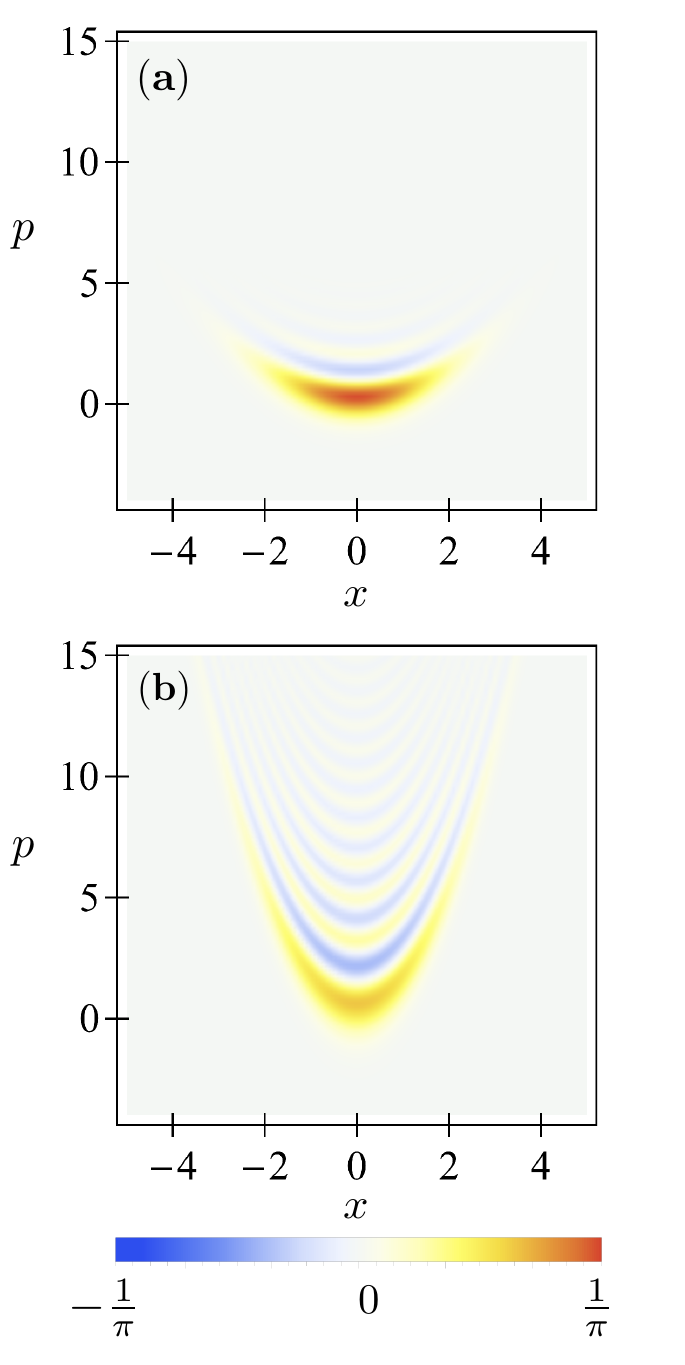}
		\caption{Wigner function $W_{\gamma,r}(x,p)$ of the state cubic phase state $\ket{\gamma,r}$ [Eq.~\Refeq{eqn:cubic-phase-state-momentum}] for ({\bf a}) cubicity $\gamma=0.1$ and squeezing $r=0.6$ ($\approx 5\dB$) and ({\bf b}) $\gamma=0.4$, $r=0.8$ ($\approx 7\dB$). 
		\label{fig:wigner-cubic-phase}}
	\end{figure}

The exact expression of the Wigner function $\nobreak{W(x,p)=\frac{1}{\pi}\int_{\mathbb{R}}\diff y\ \EXP{2ipy} \varphi(x+y)^*\varphi(x-y)}$ of the state \Refeq{eqn:cubic-phase-state-momentum} is given by
	\begin{equation}\label{WCubic}
		W_{\gamma,r}(x,p)=\mathcal{M}_{\gamma,r}\ \EXP{-\tfrac{p}{3\gamma \EXP{2r}}} \Ai\left[c\left(3\gamma x^2-p+\frac{1}{12\gamma\EXP{4r}}\right)\right] ,
	\end{equation}
	where $c=\left(\frac{4}{3\gamma}\right)^{\frac13}$, $\Ai$ is the Airy function and $\mathcal{M}_{\gamma,r}$ is a normalization factor (see \refappndx{sec:wigner-function-calculation} for details). So far the analytical expression of the Wigner distribution was known only for the unphysical cubic state $\ket{\gamma}$~\cite{ghose2007non}, while for the finite-squeezing case only implicit expressions have been used~\cite{zhuang18}. The average position and momentum are given by $\langle \hat x \rangle_{\gamma,r}=0$ and $\langle \hat p \rangle_{\gamma,r}=\frac32\gamma \EXP{2r}$, so that for increasing values of $\gamma$ and $r$ the average position remains unchanged while the momentum shifts toward positive (negative) values for positive (negative) $\gamma$. The state possesses an axial symmetry as shown in \reffig{fig:wigner-cubic-phase} and it is possible to appreciate its highly nonclassical features, as the Wigner distribution develops negative `ripples' in an extended region of the phase space.

\section{Squeezed and displaced finite superpositions of Fock states}\label{s:FiniteSup}
	In the previous Section we saw that enforcing canonical commutation relations on $\hat f$ introduces a severe constraint on the form of \refeq{fmode}, as the parameter space is reduced to a set of three real independent parameters, namely $\{r,\theta,t\}$. Given that the dark state condition \refeq{DarkState} holds true whether or not
the commutation relations are satisfied, we now look for possible unconditionally pure states of the target mode when $[\hat f,\hat f^{\dagger}]\not\propto  \mathbbm{1}$. Contrary to the case of Sec.~\ref{s:ResQuad}, their existence is in principle possible because the introduction of a linear term in the coupling breaks the discrete parity symmetry \cite{albert2014symmetries}. However, the characterization of such states is complicated by the  large parameter space.
A possible approach would consist in projecting \refeq{DarkState} onto the Fock basis and solving the resulting recurrence relation between the components of the steady state vector. Unfortunately this attempt fails, since the recurrence relation cannot be  resummed in general, at variance with the case of a linear transformation (for the derivation of the coefficients of a squeezed state in the Fock basis see, e.g.~\cite{walls2007quantum}). 

Here we pursue an alternative  approach to find analytical closed expressions for the steady state of the system: we transform \refeq{DarkState} to the {\em position} representation, where the problem is reduced to solving a differential equation for the stationary wave function. Once the wave function is known, by exploiting some properties of the Hermite polynomials it is possible to obtain the explicit expression of the steady state in the Fock basis (to find such expression we mainly used identities found in Ref.~\cite{gradshteyn2007table}). In position representation, the canonical position and momentum operators $\hat{x}=\left(b+b^\dagger\right)/\sqrt2$ and $\hat{p}=-i\left(b-b^\dagger\right)/\sqrt2$ are replaced with the position quadrature $x$ and the differential operator $\hat{p}\equiv-i\frac{\mathrm{d}}{\mathrm{d}x}$, respectively. Hence, \refeq{DarkState} becomes 
	\begin{equation}\label{eqn:diff-eqn}
		A \varphi''(x)+ B(x) \varphi'(x)+C(x) \varphi'(x)=0 \, ,
	\end{equation}
	where
	\begin{align}
		A&=\frac{c_3+c_4}{2}-c_5\, , \label{A} \\
		B(x)&=\frac{c_1-c_2}{\sqrt2}+(c_3-c_4)x\, , \label{B}\\
		C(x)&=\frac{c_3-c_4}{2}+\frac{c_1+c_2}{\sqrt2}x+\left[\frac{c_3+c_4}{2}+c_5\right]x^2. \label{C}
	\end{align}
	This is a second-order, linear, homogeneous ordinary differential equation for the steady-state wave function $\varphi(x)\equiv\bra{x}\varphi\rangle$. This equation admits at most two solutions depending on the coupling parameters $c_j$, which correspond to two pure linearly independent steady states. In contrast to the case of \refsec{s:CubicPhase}, which combination of these two gets populated in the infinite-time limit now depends on the initial state of the system. 
	Below we study in detail the only two instances of  {\em unconditionally pure states} that emerge in this scenario.


	\subsection{First family}\label{s:FirstFamily}
		The first family of unconditionally pure steady states is obtained when the quadratic couplings satisfy the following condition
		\begin{equation}\label{FirstEqCondition}
			c_5=\frac{c_3+c_4}{2}\, , \quad \mathrm{with}\quad c_3\ne\pm c_4\,. 
		\end{equation}
		Note that for $c_3 = c_4$ the case of a cubic phase state is recovered, while for $c_3=-c_4$ the state is not normalised. Under these assumptions, the coefficient of the second derivative \refeq{A} identically vanishes and the steady-state differential equation \Refeq{eqn:diff-eqn} reduces to a first order equation, whose solution is given by
		\begin{equation}\label{eqn:solution}
			\varphi(x)\propto\ (x+x_1)^\epsilon\ \EXP{-\alpha(x-z_1)^2}\ ,
		\end{equation}
		with the expressions of the complex coefficients $x_1$, $z_1$, $\alpha$ and $\epsilon$  given in \refappndx{sec:expressions}.

		The solution \Refeq{eqn:solution} comes into the form of a Gaussian function times a power law function, and hence it is not physical, i.e., square-integrable, for all values of the parameters $c_j$. However, provided that the conditions ensuring the physicality of the solution are met, it describes unique and pure steady state. We restrict our study to the case of real parameters  $x_1,z_1,\alpha\in\mathbb{R}$ and impose $\epsilon$ to be a natural number, i.e., $\epsilon\stackrel{!}{=}n\in\mathbb{N}$. In this case we write the wave function as $\varphi(x)\equiv\varphi_n(x)$, which turns out to be square-integrable whenever $\alpha>0$.  The introduction of an integer parameter represents a key feature of our approach: it sets a  constraint among the coefficients [see \refeq{epsilon}] that is crucial to obtain a simple analytical expression for the steady state, without at the same time imposing a prohibitive condition for the physical implementation of the model. Indeed, in a cavity optomechanics setup, such integer condition is simply achieved by tuning the relative strength among the control lasers that drive the optical cavity (see  \refsec{s:Optomech}). Furthermore, in \refsec{s:Imprecision} we provide numerical evidence that this condition can be realized also with finite accuracy.

		In order to obtain the explicit expression of the state from the wave function $\varphi_n(x)$, we write \refeq{eqn:solution} in terms of the eigenstates of the harmonic oscillator. By changing variable to $y=\frac{1}{\sqrt{2\alpha}}(x-z_1)$, the wave function becomes
		\begin{align}
			\Phi_n(y)	&\propto	 (y+\lambda)^n\EXP{-\frac12 y^2}\, ,\nonumber\label{eqn:family-1-wave-function}\\
						&=	 \sum\limits_{k=0}^n\binom{n}{k}\ \lambda^{n-k}y^k \EXP{-\frac12 y^2}\ ,
		\end{align}
		where we set $\lambda=\sqrt{2\alpha}(x_1+z_1)$. Next, the monomials $y^k$ can be written in terms of Hermite polynomials $\Hermite_{m}(y)$ as~\cite{gradshteyn2007table}
		\begin{equation}
			y^k=\sum\limits_{m=0}^{\lfloor \frac{k}{2}\rfloor}\frac{k!}{2^k m!(k-2m)!}\ \Hermite_{k-2m}(y)\,,
		\end{equation}
		where $\lfloor x \rfloor$ yields the greatest integer smaller or equal than $x$.
		Consequently, the wave function expression becomes
		\begin{equation}
			\Phi_n(y)\propto\sum\limits_{k=0}^n\sum\limits_{m=0}^{\lfloor \frac{k}{2}\rfloor}\binom{n}{k}\frac{k! \lambda^{n-k}}{2^k m! (k-2m)!}\ \Hermite_{k-2m}(y) \EXP{-\frac12 y^2}\,.
		\end{equation}
		Notice that the last two factors in this expression are related to the wave function of the harmonic oscillator, $\bra{y}n\rangle=\frac{\pi^{-1/4}}{\sqrt{2^n n!}}\ \EXP{-\frac12 y^2}\ \Hermite_n(y)$. Therefore, the corresponding state vector in Fock basis has the following form
		\begin{equation}\label{eqn:family-1}
			\ket{\Phi_n}=\mathcal{N}_{\Phi_n}\sum\limits_{k=0}^n\sum\limits_{m=0}^{\lfloor \frac{k}{2}\rfloor}\binom{n}{k}\frac{\pi^{\frac14}k! \lambda^{n-k}}{2^{\frac{k}{2}+m} m! \sqrt{(k-2m)!}}\ \ket{k-2m},
		\end{equation}
		where $\mathcal{N}_{\Phi_n}$ is a normalisation factor (see \refappndx{sec:expressions}).
		Finally, the change of variable $y=\frac{1}{\sqrt{2\alpha}}(x-z_1)$ done above corresponds, in Fock space, to a displacement and a squeezing operation. Therefore, taking these two operations into account, the steady state of the target mode reads
		\begin{eqnarray}\label{eqn:DS-family-1}
			\ket{\varphi_n}	&=&	\displ\bigl(z_1/\sqrt2\bigr)\sqz \bigl(\ln \sqrt{2\alpha}\bigr) \ket{\Phi_n},
		\end{eqnarray}
		where \displ\ is the displacement operator defined as $\nobreak{\displ(\beta)=\EXP{\beta\hat b^\dagger-\beta^*\hat b}}$.

		The expression \Refeq{eqn:DS-family-1} [or equivalently \Refeq{eqn:family-1}] characterizes a novel class of states of a bosonic system. The most important feature of this class is that, apart from the two Gaussian operations, it consists of a  {\em finite superposition} of Fock states. Moreover, each $\ket{\Phi_n}$ contains at most $n$ excitations, so that the integer parameter -- initially introduced for the sake of convenience -- acquires a well-defined physical meaning. All states (except the case $n=0$) are non-Gaussian and thus by Hudson's theorem they all have negative Wigner function~\cite{hudson1974wigner}. 
		
		Past theoretical works proposed probabilistic methods for the truncation of photon number superpositions in linear optical systems~\cite{pegg1998optical,ozdemir2001quantum,paul1996photon}, which however require multiple post-selections and are plagued by low efficiency. Protocols to prepare arbitrary finite superpositions of  travelling photons~\cite{dakna1999generation,fiuravsek2005conditional} or cavity photons interacting with atomic probes~\cite{vogel1993quantum} have been also put forward, but they either require iterative measurements and post-selection or multiple nonlinear operations such as single-photon addition.
		In contrast, here a finite superposition of a \emph{desired number of elements} is obtained unconditionally, which is a unique feature of our approach~\footnote{The state Eq.~\Refeq{eqn:DS-family-1} can always be deterministically counter-squeezed and displaced to isolate the finite superposition $\ket{\Phi_n}$.}. As is clear from the expression, the superposition Eq.~\Refeq{eqn:DS-family-1} takes a specific form and its coefficients cannot be arbitrarily tuned. 
	However, this family allows enough flexibility to stabilize a variety of interesting quantum states, as will be shown in Sec.~\ref{s:ExFiniteSup}.  

		Since the states under scrutiny are displaced and squeezed finite superpositions of number states, any negativity of the Wigner function will not result from the action of the Gaussian operation, and the superposition of number states must be the sole source of negativity of the Wigner function. Therefore, we focus on Wigner function of the superposition. The Wigner function associated to the states \Refeq{eqn:family-1} is given by 
		\begin{widetext}
			\begin{equation}\label{WPhi}
				W_{\Phi_n}(x,p)=\mathcal{M}_{\Phi_n}\, \EXP{-(x^2+p^2)}\sum\limits_{k=0}^n\ (-1)^k\frac{(x+\lambda)^{2(n-k)}}{(n-k)!}\ \mathrm{L}_k^{(-\frac12)}(p^2)\, ,
			\end{equation}
		\end{widetext}
		where the normalization factor $\mathcal{M}_{\Phi_n}$ and the expression of $\lambda$ are given in \refappndx{sec:wigner-function-calculation} and $\mathrm{L}_n^{(\nu)}(y)$ are the generalized Laguerre polynomials.

	\subsection{Second family}\label{s:SecondFamily}
		A second family of unconditionally pure states is obtained by letting the coupling parameters satisfy $c_5^2\ne c_3 c_4$ and $c_5\ne\frac12(c_3+c_4)$. In this case, the two linearly independent solutions of the steady-state differential equation \Refeq{eqn:diff-eqn} are 
		\begin{eqnarray}
			\varphi_1(x)	&\propto&	\EXP{-\alpha'(x-x_2)^2}\ \Hermite_\eta\left(\tau(x-z_2)\right)\label{eqn:solution-1}\\
			\varphi_2(x)	&\propto&	\EXP{-\alpha'(x-x_2)^2}\ _1\mathrm{F}_1(-\eta,\frac12,\tau^2(x-z_2)^2)\ ,\label{eqn:solution-2}
		\end{eqnarray}
		where $_1\mathrm{F}_1$ is the confluent hypergeometric function  and $x_2$, $z_2$, $\alpha'$, $\eta$ and $\tau$ are complex coefficients reported in \refappndx{sec:expressions}. While the solution Eq.~\Refeq{eqn:solution-1} is normalizable for some parameters $g_j$, Eq.~\Refeq{eqn:solution-2} is never square integrable. Therefore also in this case the system admits an unconditionally pure solution. Again, we restrict our study to the case of real parameters $x_2,z_2,\alpha',\tau\in\mathbb{R}$ and impose  $\eta\stackrel{!}{=} n$, in which case a sufficient condition of physicality of the solution \Refeq{eqn:solution-1} is $\alpha'>0$. In the following we put $\varphi_1\equiv\psi_n$.

		As before, we write the wave function Eq.~\Refeq{eqn:solution-1} in terms of the eigenstates of the harmonic oscillator. In fact, by putting $y=\sqrt{2\alpha'}\ (x-z_2)$, $s=\frac{\tau}{\sqrt{2\alpha'}}$ and $u=\frac{1}{\sqrt{2\alpha'}}(x_2-z_2)$, the wave function transforms to 
		\begin{eqnarray}
			\Psi_n	&\propto&	 \EXP{-\frac12 y^2}\ \Hermite_n\bigl(s(y-u)\bigr)\ ,\label{eqn:family-2-wave-function}\\
					&=&	 \EXP{-\frac12 y^2}\ \sum\limits_{k=0}^{\lfloor\frac{n}{2}\rfloor}\ \frac{n!}{k!(n-2k)!}\ s^{n-2k}(s^2-1)^k\ \Hermite_{n-2k}(y-u)\ ,\nonumber\\
					&=&	 \EXP{-\frac12 y^2}\ \sum\limits_{k=0}^{\lfloor\frac{n}{2}\rfloor}\ \frac{n!}{k!(n-2k)!}\ s^{n-2k}(s^2-1)^k\ \nonumber \\
					&\times& \sum\limits_{m=0}^{n-2k}\ \binom{n-2k}{m}(-2u)^{n-2k-m}\ \Hermite_m(y)\ ,
		\end{eqnarray}
		from which we obtain the expression of the state
		\begin{align}\label{eqn:family-2}
			\ket{\Psi_n}&=\mathcal{N}_{\Psi_n}\sum\limits_{k=0}^{\lfloor\frac{n}{2}\rfloor} \sum\limits_{m=0}^{n-2k}
			\frac{\pi^{\frac14}2^{\frac{m}{2}}n!}{k! \sqrt{m!} (n-2k-m)!} \nonumber \\
			&\times s^{n-2k}(s^2-1)^k (-2u)^{n-2k-m}\ket{m}\ ,
		\end{align}
		with $\mathcal{N}_{\Psi_n}$ a normalization factor given in \refappndx{sec:expressions}. The steady state $\ket{\psi_n}$ is thus given by
		\begin{equation}\label{eqn:DS-family-2}
			\ket{\psi_n}=\displ\bigl(z_2/\sqrt2\bigr)\sqz \bigl(\ln \sqrt{2\alpha'}\bigr) \ket{\Psi_n}\, ,
		\end{equation}
		
		while the Wigner function of the finite superposition $\ket{\Psi_n}$ is equal to
		\begin{widetext}
		\begin{equation}\label{WPsi}
			W_{\Psi_n}(x,p)=\mathcal{M}_{\Psi_n}\, \EXP{-x^2-p^2}\sum\limits_{k=0}^n\ \binom{n}{k}^2 k!(-2s^2)^k(1-s^2)^{n-k}\left|\Hermite_{n-k}\left(\frac{s}{\sqrt{1-s^2}}(x-u+ip)\right)\right|^2\, ,
		\end{equation}
		\end{widetext}
		with $\mathcal{M}_{\Psi_n}$ a normalisation constant (see \refappndx{sec:wigner-function-calculation}).


\section{Examples of finite superposition}\label{s:ExFiniteSup}
	
	After the exhaustive study provided in the previous Section, to uncover the full potential of linear-and-quadratic reservoir engineering we now focus on some relevant examples of finite superpositions. Indeed, within families ~\eqref{eqn:family-1} and~\eqref{eqn:family-2} are instances of
	states with a particularly simple expression and whose features resemble those of some of the most paradigmatic states of the quantum harmonic oscillator, such as the Sch\"odinger cat state and Fock states.
	
	\begin{figure*}[t!]
		\centering
		\includegraphics[scale=.85]{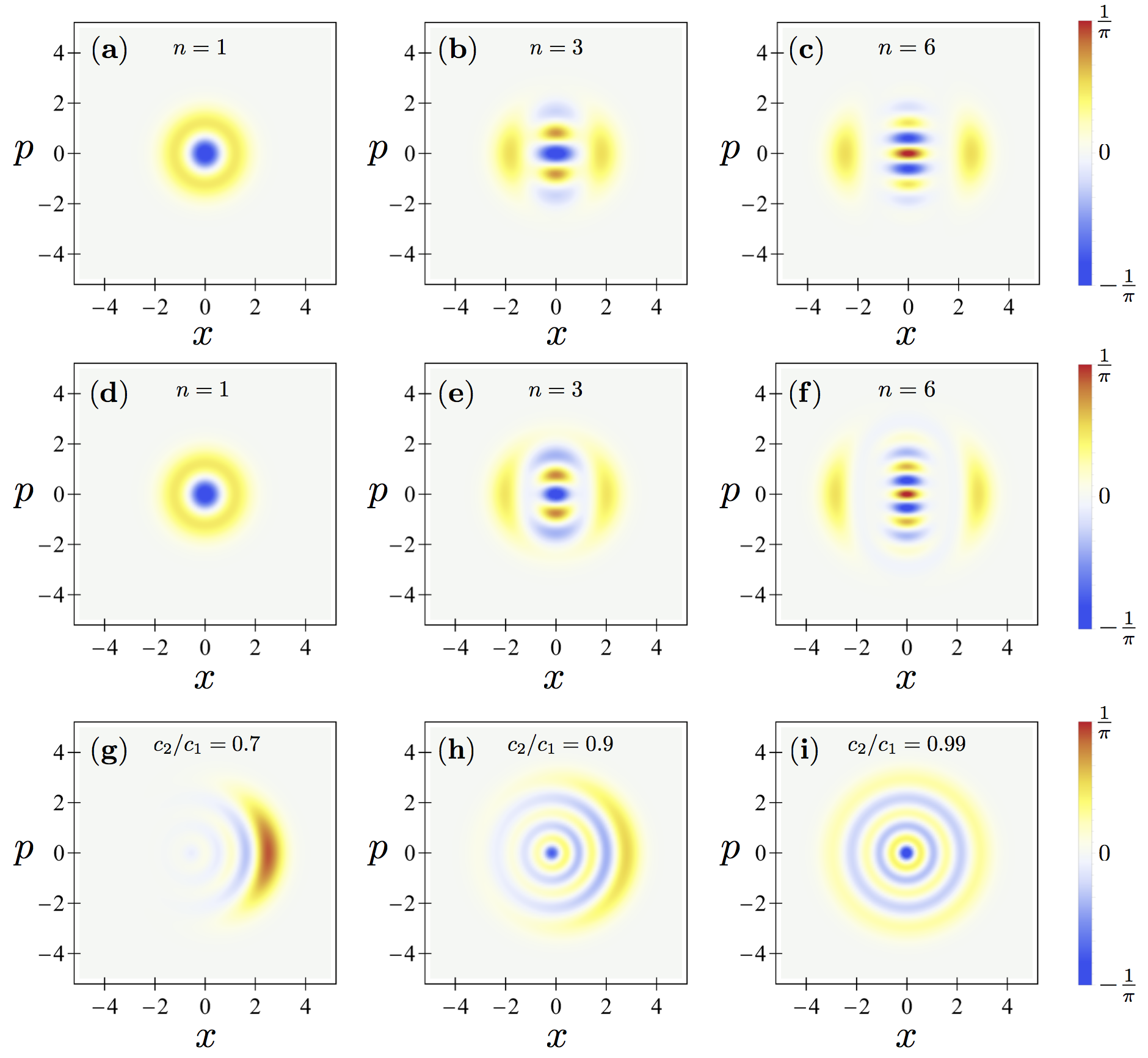}
		\caption{Relevant examples of finite superpositions of Fock states. ({\bf a})-({\bf c}) Wigner function of the state $\ket{\mathcal{C}^\Phi_n}$ [Eq.~\eqref{FirstCatLike}] for $n=1,3,6$;  ({\bf d})-({\bf f})
		 Wigner function of the state $\ket{\mathcal{C}^\Psi_n}$ [Eq.~\eqref{SecondCatLike}] for $n=1,3,6$;  ({\bf g})-({\bf i}) Wigner function of the state
		 $\ket{\mathcal{F}^\Psi_n}$ [Eq.~\eqref{FockLike}] with $n=5$ for $c_2/c_1=0.7,0.9,0.99$  ($\approx 7,13,23\dB$).
		\label{f:PlotSuperposition}}
	\end{figure*}


	\subsection{Two new families of macroscopic quantum superposition}
		Expression~\eqref{eqn:family-1} greatly simplifies if we impose $\lambda\stackrel{!}{=}0$. In this way, only the terms with $k=n$ will have a non-zero contribution to the sum
		and the state $\ket{\Phi_n}$ reduces to the following simple expression 
		\begin{equation}\label{FirstCatLike}
			\ket{\mathcal{C}^\Phi_n}\propto\sum\limits_{m=0}^{\lfloor \frac{n}{2}\rfloor}\frac{1}{2^{m} m! \sqrt{(n-2m)!}}\ \ket{n-2m}.
		\end{equation}
		Notice that this state is a superposition of $\lfloor \frac{n}{2}\rfloor+1$ Fock states, ranging from $\ket{0}$ to $\ket{n}$, and has definite number parity depending on whether $n$ is even/odd. In Fig.~\ref{f:PlotSuperposition} ({\bf a})-({\bf c}) we plot the Wigner distribution of such state for different values of $n$. For $n=1$ the state 
		contains a single excitation, i.e.,~$\ket{\mathcal{C}^\Phi_1}\equiv\ket{1}$, while for increasing  $n$ we clearly see that it acquires the distinctive features of a macroscopic quantum superposition. The corresponding dark state state Eq.~\Refeq{eqn:DS-family-1} reads $\ket{\varphi_n}=\displ\bigl(-\sqrt{2n+1}\bigr)\sqz \bigl(-\ln \sqrt{2}\bigr) \ket{\mathcal{C}^\Phi_n}$ and is the unique state annihilated by the nonlinear operator
		\begin{equation}\label{FirstCatMode}
			\hat f=c_1 \hat b+\frac{c_1}{4\sqrt{2(2n+1)}} \left( 3\hat b^2- \hat b^{{\dag}\,2}+\{\hat b^{\dag} ,\hat b \}\right) \, .
		\end{equation} 
		This expression can be obtained from the constraints $\lambda=0$, $\epsilon=n$, together with Eq.~\eqref{FirstEqCondition},
		and provides the universal set of coefficients $c_j$ to be imposed for the stabilization of the state.
		Eq.~\eqref{FirstCatMode} is a quite remarkable result because, despite its simplicity, it would have been difficult to predict by other means---in particular guided by symmetry arguments alone.
		The detour via the wave function Eq.~\eqref{eqn:diff-eqn} makes instead possible to uncover this peculiar linear-and-quadratic combination of operators and determine the unique state  annihilated by it.
		
\begin{figure}[tb]
		\centering
		\includegraphics[width=\columnwidth]{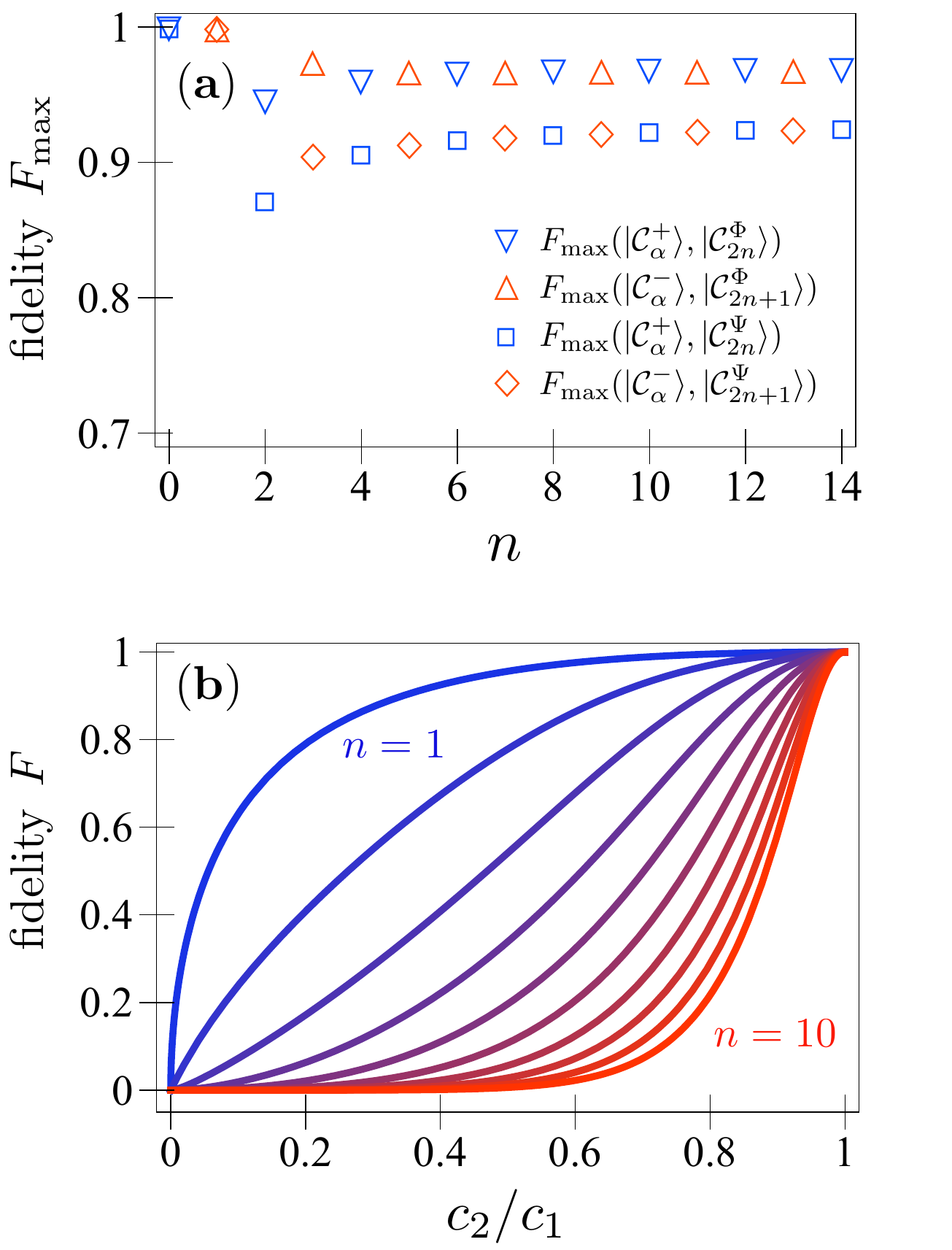}
		\caption{({\bf a}) Maximum fidelity between the macroscopic quantum superpositions $\ket{\mathcal{C}^\Phi_n}$ [Eq.~\eqref{FirstCatLike}], $\ket{\mathcal{C}^\Psi_n}$ [Eq.~\eqref{SecondCatLike}] and the Schr\"odinger cat state $\ket{\mathcal{C}_{\alpha}^\pm}\propto\ket{\alpha}\pm\ket{-\alpha}$ for different values of $n$; the fidelity is optimized over $\alpha$.
		({\bf b}) Fidelity between $\ket{\mathcal{F}^\Psi_n}$ [Eq.~\eqref{FockLike}] and the Fock state $\ket{n}$ as a function of the amount of squeezing, from $n=1$ (blue)
		to $n=10$ (red).
		\label{f:PlotFidelity}}
	\end{figure}



		Along the same lines, we can simplify Eq.~\eqref{eqn:family-2}  by imposing $u\stackrel{!}{=}0$, in such a way that only the terms with $m=n-2k$ will contribute to the expression. This condition leads to 
		the state 
		\begin{equation}\label{SecondCatLike}
			\ket{\mathcal{C}^\Psi_n}\propto \sum\limits_{m=0}^{\lfloor \frac{n}{2}\rfloor}\frac{1}{4^{m} m! \sqrt{(n-2m)!}}\ \ket{n-2m},
		\end{equation}
		which for $n=1$ reduces to $\ket{\mathcal{C}^\Psi_1}\equiv\ket{1}$ and for larger $n$ represents a second family of macroscopic quantum superpositions. Notice the similarity between the expressions Eqs.~\eqref{SecondCatLike} and~\eqref{FirstCatLike}, which  have the same parity and span and differ just for a term in the coefficient. In Fig.~\ref{f:PlotSuperposition} ({\bf d})-({\bf f}) we show the plots of the corresponding Wigner distribution.  The nonlinear operator that annihilates the state $\nobreak{\ket{\psi_n}=\displ\bigl(-\sqrt{2n+1}\bigr)\sqz \bigl(-\ln \sqrt{3}\bigr) \ket{\mathcal{C}^\Psi_n}}$ is given by
		\begin{equation}\label{SecondCatMode}
			\hat f=c_1 \hat b-\frac{c_1}{2\sqrt{2n+1}} \left( \hat b^{{\dag}\,2}-\{\hat b^{\dag} ,\hat b \}\right) 
		\end{equation}
		and involves one term less than Eq.~\eqref{FirstCatMode}.
		This case of macroscopic superposition was studied in detail in Ref.~\cite{brunelli2018unconditional}. The expressions Eqs.~\eqref{FirstCatMode},~\eqref{SecondCatMode} share a similar structure, namely that of  a cooling process coherently superimposed to a set of nonlinear operators; larger superpositions are 
		 obtained for larger $n$ [see Fig.~\ref{f:PlotSuperposition} ({\bf c}), ({\bf f})], which in turn entails weaker nonlinear contributions.



A quantitative assessment of the `macroscopicity' of the target states $\ket{\mathcal{C}^{\Phi,\Psi}_n}$ can be obtained by comparing  them 
with a Schr\"{o}dinger cat state, which provides the benchmark for macroscopic  superposition states. We consider the cat states  
$\nobreak{\ket{\mathcal{C}_{\alpha}^\pm}=\bigl[2\bigl(1\pm e^{-2\vert\alpha\vert^2}\bigr)\bigr]^{-\frac12}(\ket{\alpha}\pm\ket{-\alpha})}$,
where the plus (minus) sign selects an even (odd) 
cat, namely a superposition of only even (odd) number states. The fidelity with the two states of Eqs.~\eqref{FirstCatLike} and ~\eqref{SecondCatLike} is given by
 $\nobreak{F^\pm_{\Phi,\Psi}(\alpha,n)=\vert 
\bigl\langle\mathcal{C}_{\alpha}^\pm\ket{\mathcal{C}^{\Phi,\Psi}_n}\vert}$ and, since both states have definite parity, the only nonzero overlaps 
are between an even (odd) cat state and an even (odd) finite superposition.
In Fig.~\ref{f:PlotFidelity} ({\bf a}) we show the maximum fidelity 
$F^{\pm}_{\mathrm{max}}=F^{\pm}_{\Phi,\Psi}(\alpha_{\mathrm{max}},n)$, optimized over 
$\alpha$; the fidelity
always lies within the range $F^{\pm}_{\mathrm{max}} \approx 0.9-1$ and, by increasing $n$, it saturates to a value 
$F^{\pm}_{\mathrm{max}}\approx 0.97$ for $\ket{\mathcal{C}^{\Phi}_n}$ and $F^{\pm}_{\mathrm{max}}\approx 0.92$ 
for $\ket{\mathcal{C}^{\Psi}_n}$. The optimal values of $\alpha$ (not shown) correspondingly 
increase, witnessing  the increasing in the macroscopic character of the superposition. 
However, the fact that the fidelity does not approach 1 provides further evidence  that 
$\ket{\mathcal{C}^{\Phi}_n}$ and $\ket{\mathcal{C}^{\Psi}_n}$ are similar but  {\it distinct} instances 
of macroscopic superposition with respect to the celebrated Schr\"{o}dinger cat state.


	\subsection{Approximated Fock state of any order}
		By inspecting Eq.~\eqref{eqn:family-2} we see that a second simple instance is obtained when we impose $s^2\stackrel{!}{=}1$. In this way only $k=0$ terms contribute to the
	expression, that becomes
\begin{equation}\label{FockLike}
\ket{\mathcal{F}_n^{\Psi}}\propto \sum_{k=0}^n \binom{n}{k}d_n^{\,-k} \ket{k}\, ,
\end{equation}
where the coefficients are $d_n=\frac{(1-\zeta)}{4\zeta}\sqrt{\zeta(2n+1)}$ and we set $\zeta=\tanh r\equiv \frac{\vert c_2\vert}{\vert c_1\vert}$.  
The corresponding state Eq.~\eqref{eqn:DS-family-2} reads $\nobreak{\ket{\psi_n}=\displ\bigl(-\sqrt{\zeta(n+1/2)}\,\bigr) \ket{\mathcal{F}^\Psi_n}}$ and is annihilated by the nonlinear operator
\begin{equation}\label{FockMode}
		\hat f=\mathcal{G}\hat \beta+\sqrt{\frac{\cosh r\sinh r}{2(2n+1)}} \{\hat b^{\dag} ,\hat b \} \, .
	\end{equation} 
	The normalized expressions of the states  Eqs.~\eqref{FirstCatLike},~\eqref{SecondCatLike} and~\eqref{FockLike} are 
given  in \refappndx{sec:expressions}.
In Fig.~\ref{f:PlotSuperposition} ({\bf g})-({\bf i}) we show the density profile of the Wigner function for different values of the squeezing parameter $\zeta$
and the same $n$ ($n=5$). We clearly see that the distribution, which for lower values of the squeezing is skewed toward one side, progressively straightens to approach that of a Fock state. The state $\ket{\mathcal{F}^\Psi_n}$ is a superposition of $n+1$ number states that displays Fock-like features, namely that `mimics' the Fock state $\ket{n}$, to an extent that improves with the amount of available squeezing. Indeed, it is easily checked that in the limit $\zeta\rightarrow1$ the superposition $\ket{\mathcal{F}^\Psi_n}$
collapses to the single element $\ket{n}$. However, this limit corresponds to an infinite amount of squeezing and  
a dynamical instability in Eq.~\eqref{SteadyL} is encountered in this case. Yet, it is still possible to approximate with near-unit fidelity any Fock state 
for finite squeezing. This feature is explored quantitatively in Fig.~\ref{f:PlotFidelity} ({\bf b}), where we show the fidelity between the state $\ket{\mathcal{F}_n^{\Psi}}$ and the Fock state $\ket{n}$ as a function of the squeezing for
different values of $n$. Not surprisingly, when increasing the number state $\ket{n}$ that we want to approximate,  achieving a given threshold fidelity requires increasing amount of squeezing. 

We finally point out that, although in this Section we focused on `minimal' cases with a concise expression and a symmetric phase space distribution, 
many other interesting instances of Eqs.~\eqref{eqn:family-1} and~\eqref{eqn:family-2} are in fact possible for suitable choices of the coupling parameters $c_j$, all featuring non-classical
behavior such as asymmetric peaks and dips in phase space.
%

%


\section{Effects of thermal environment and imprecisions}\label{s:Imprecision}
\begin{figure*}[t!]
		\centering
		\includegraphics[scale=0.75]{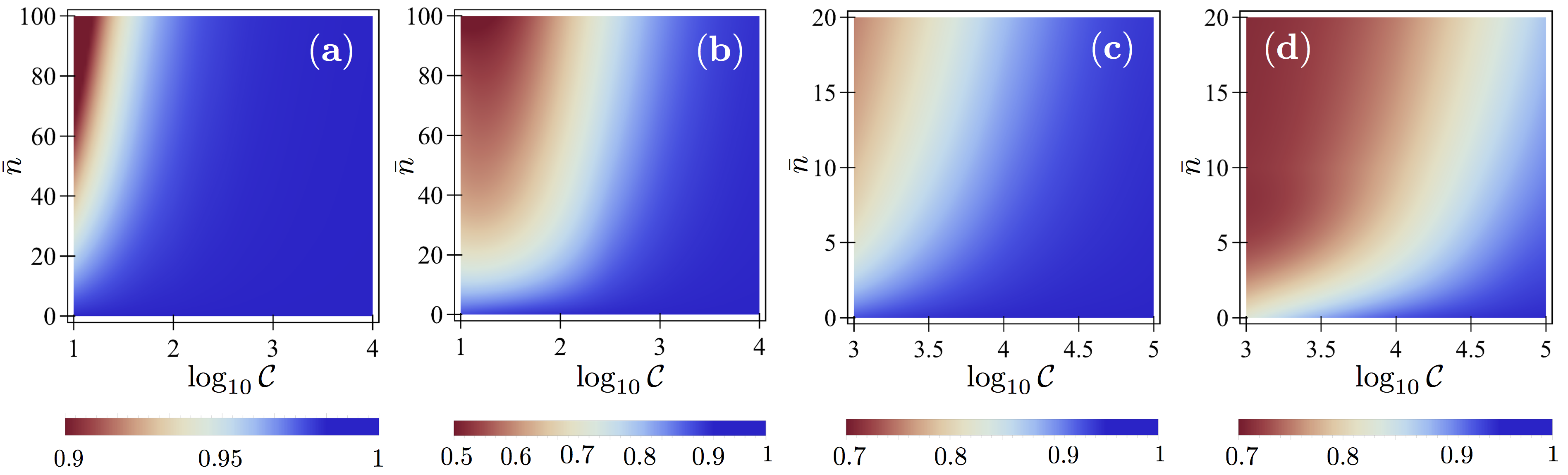}
		\caption{Fidelity between the mechanical steady state in the presence of thermal noise and the ideal target state [$\gamma_m=0$ in Eq.~\eqref{Adiabatic}]
		as a function of $\bar n$ and $\gamma_m$ (parametrized by the cooperativity $\mathcal{C}$).
		({\bf a}) Cubic phase state [Eq.~\eqref{eqn:cubic-phase-state-momentum}] with $\gamma=0.1$, $r=0.6$. 
		({\bf b}) Displaced Fock-like state [see Eq.~\eqref{FockLike} and following paragraph] for $n=5$, $r=0.69$. ({\bf c}), ({\bf d}) Displaced and squeezed finite superposition
		of Fock states [see Eq.~\eqref{SecondCatLike} and following paragraph] for $n=3$ ({\bf c}) and $n=5$ ({\bf d}).\label{f:PlotFidelityTh} }
	\end{figure*}
We now address how the unavoidable presence of dissipation affects the properties of the target state.
To this aim, we assume that the target mode $\hat b$ is in contact with a thermal environment  at finite temperature, so that the 
overall evolution  is modified as follows 
\begin{equation}
\dot{\hat \varrho}=\mathcal{L}[\hat \varrho]+
\gamma_m (\bar n+1)\mathcal{D}_{ b}[\hat \varrho]+\gamma_m \bar n \mathcal{D}_{ b^\dag}[\hat \varrho] \, ,
\end{equation}
where $\gamma_m$ is the dissipation rate,  $\bar n$ the thermal occupancy and $\mathcal{L}[\hat \varrho]$ 
includes the coherent interaction and the dissipation on the auxiliary mode, as per Eq.~\eqref{MasterEq}. For concreteness, 
already in this Section we will adopt the optomechanical terminology, anticipating the discussion of Sec.~\ref{s:Optomech}; we will thus refer to the target (auxiliary) mode as the mechanical (cavity) mode. However, we remind that the conclusions hold in  general for linear-and-quadratic coupling between two modes. 		 
 For simplicity, we focus on the fast cavity limit 
$\kappa\gg c_j$, where adiabatic elimination of the cavity field leads to an effective master equation for the  reduced 
mechanical density matrix~\cite{gardiner2004quantum}
\begin{align}\label{Adiabatic}
\dot{\hat \varrho}^{(m)}&=\gamma_m \mathcal{C}\,\mathcal{D}_{ f/\mathcal{G}}\,\bigl[\hat \varrho^{(m)}\bigr] \nonumber \\
&+\gamma_m (\bar n+1)\mathcal{D}_{ b}\bigl[\hat \varrho^{(m)}\bigr]+\gamma_m \bar n \mathcal{D}_{ b^\dag}\bigl[\hat \varrho^{(m)}\bigr] \,,
\end{align}
where $\mathcal{C}=4\mathcal{G}^2/(\gamma_m\kappa)$ quantifies the cooperativity. 
The first term on the right-hand 
side describes an effective dissipation induced by the modified jump operator $\hat f/\mathcal{G}$; the
second and third term describe loss and incoherent pumping to/from the environment.
The appearance of the term  $\mathcal{D}_{ f/\mathcal{G}}$ makes manifest the action of the auxiliary mode as an engineered reservoir for the target mode.
We also stress that the jump operator of such an effective dissipator is nonlinear and non-bosonic.
\par
\par
We consider some of the relevant steady states encountered in the previous Sections and study the impact of 
the thermal environment on them. 
We numerically find the steady state $\hat \varrho^{(m)}_{ss}$ of \refeq{Adiabatic} for different choices of $\hat f$ and compare it to the
ideal steady state obtained in the absence of noise \footnote{We used QuTiP \cite{johansson2013qutip,johansson2012qutip} to calculate the steady state of \refeq{Adiabatic}}. Fig.~\ref{f:PlotFidelityTh} ({\bf a}) shows the fidelity between $\hat \varrho^{(m)}_{ss}$ and 
the ideal cubic phase state of  Eq.~\eqref{eqn:cubic-phase-state-momentum} when $\hat f$ is nonlinear Bogoliubov transformation; 
Fig.~\ref{f:PlotFidelityTh} ({\bf b}) the fidelity with the Fock-like state $\nobreak{\ket{\psi_n}=\displ\bigl(-\sqrt{\zeta(n+1/2)}\,\bigr) \ket{\mathcal{F}^\Psi_n}}$ 
for $\hat f$ as in Eq.~\eqref{FockMode}. Finally, Figs.~\ref{f:PlotFidelityTh} ({\bf c}) and ({\bf d}) show the fidelity between the macroscopic quantum superposition 
$\nobreak{\ket{\psi_n}=\displ\bigl(-\sqrt{2n+1}\bigr)\sqz \bigl(-\ln \sqrt{3}\bigr) \ket{\mathcal{C}^\Psi_n}}$ and the steady state with modified jump operator as in 
Eq.~\eqref{SecondCatMode} and $n=3$, $n=5$, respectively.
 As expected, mechanical dissipation is responsible for a decrease of the purity of the steady state, and therefore for non-unit fidelity with the 
 ideal target state. Of the examples shown, the cubic phase state is the most robust with respect to thermal noise, while displaced and squeezed finite 
 superpositions are more affected. In particular, we notice how the family of cat-like states $\ket{\psi_n}$  [panels ({\bf c}) and ({\bf d})] is particularly 
 susceptible to thermal decoherence, and regions of near-unit fidelity reduce considerably by increasing $n$, i.e., increasing the extent of the superposition;
 this behavior  reflects the fragility of macroscopic quantum superpositions to environmental noise. 
Nevertheless, we see that regions of near-unit fidelity are present even for non-zero thermal occupancy. Moreover, even when the fidelity is no longer close to 
one, it can be show that the steady state  $\hat \varrho^{(m)}_{ss}$ is still non-Gaussian and retains negative portions of the Wigner function for a wide range 
of parameters (see Ref.~\cite{brunelli2018unconditional} for details).  
		
\begin{figure}[t!]
		\centering
		\includegraphics[width=\columnwidth]{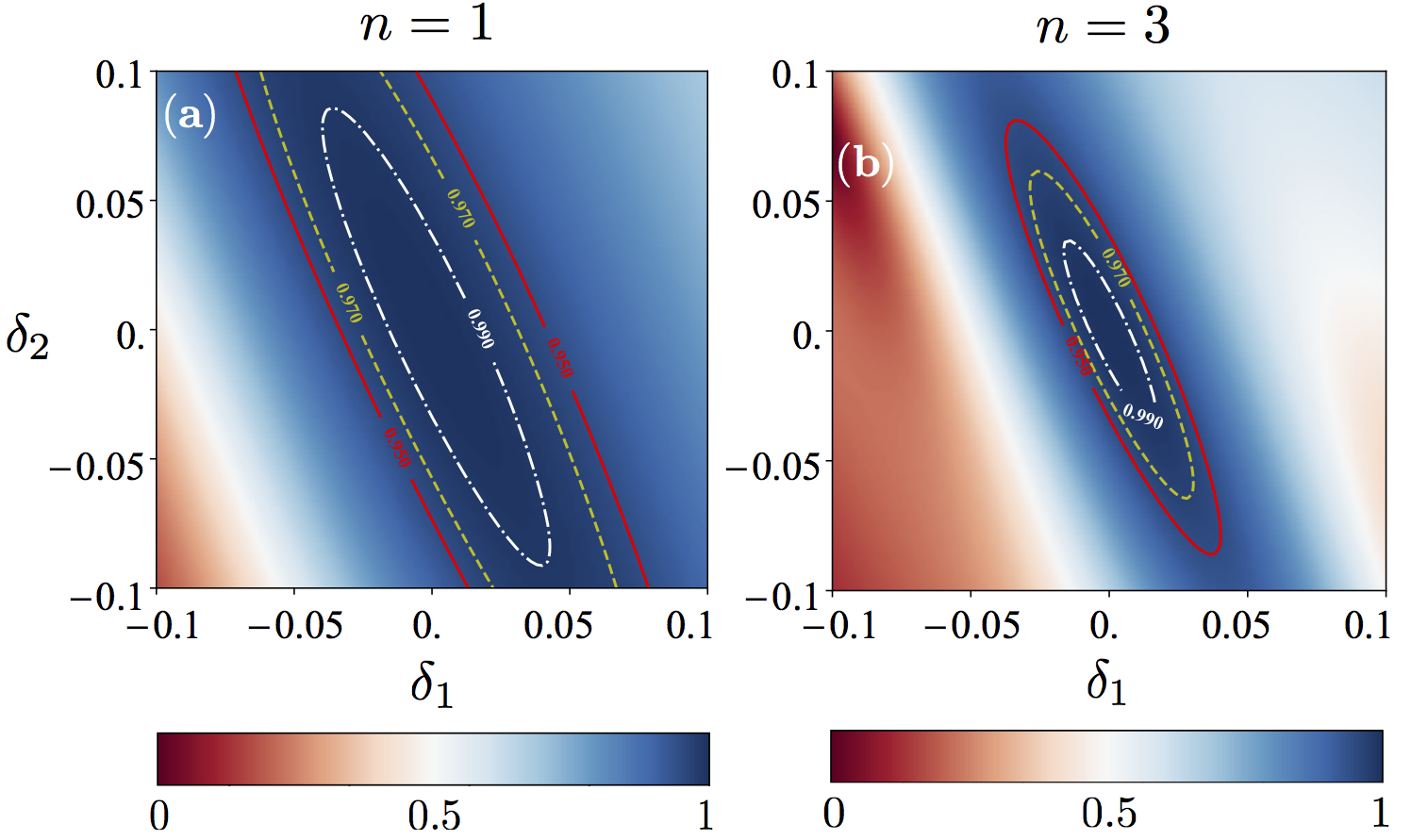}
		\caption{Fidelity between the target state annihilated by Eq.~\eqref{SecondCatMode} and the steady state obtained with perturbed couplings. 
		The fidelity is plotted against relative errors $\delta_1$ ($\delta_2$) in the coupling with the term $\hat b^{\dagger\,2}$ ($\{ \hat b,\hat b^{\dagger}\}$) for $n=1$ ({\bf a}) and $n=3$ ({\bf b}). In the above plots, the shown contours correspond to fidelities of 0.95 (red, solid), 0.97 (yellow, dashed) and 0.99 (white, dot-dashed).\label{f:PlotImprecision}}
	\end{figure}

		A second source of imprecision comes from the finite accuracy in tuning the coefficients to the desired ratios. 
		Indeed, in order to obtain the desired steady states, some (or all) of the coupling coefficients 
		$c_j$, $j=1,\ldots,5$ have to take specific values. As detailed in the next Section,
		in an optomechanical implementation this constraint  translates into tuning the amplitude and phase of the driving fields, which cannot be done 
		with arbitrary precision. 
		Therefore we study the effects of small imprecision in tuning the driving fields to the desired ratios. For concreteness, we focus on the instance 
		 $\nobreak{\ket{\psi_n}=\displ\bigl(-\sqrt{2n+1}\bigr)\sqz \bigl(-\ln \sqrt{3}\bigr) \ket{\mathcal{C}^\Psi_n}}$, but similar conclusions hold for all the other cases. We 
		model such imprecisions by adding small deviations to the exact values of the couplings shown in Eq.~\eqref{SecondCatLike}; we add independent offsets to the          
		 second and third nonlinear term ($c_4$ and $c_5$) and consider the relative error with respect to the ideal value. In Fig.~\ref{f:PlotImprecision} we show the fidelity of the steady state with respect to the ideal one
		as a function of these relative errors. We can see that accuracy within few percent is enough to guarantee near-unit values of the fidelity. 
		
Finally, we stress that our reservoir engineering approach relies on having two independent dissipation channels acting on the target and the auxiliary mode. While this 
assumption (which stands behind the local master equation approach) is well justified in the weak coupling limit we are interested in, one in general needs to be careful~\cite{Breuer:03}. In particular, away from the weak coupling limit the optical and mechanical modes hybridize, leading to nonlocal dissipators, which would hinder the effectiveness of our method. This issue becomes particularly relevant when addressing thermodynamic considerations, which however are not the focus of our work~\cite{Naseem18}.  

\section{optomechanical implementation}\label{s:Optomech} 
	In this section we show how the abstract model studied above can be realized in a cavity optomechanical setup. We consider an optomechanical system where the frequency of a cavity mode parametrically couples to the displacement and  the displacement squared of a mechanical resonator~\cite{aspelmeyer2014cavity}. The total Hamiltonian is given by 
	\begin{align}
		\hat H&=\hat H_0+\hat H_{\text{int}}+\hat H_{\text{drive}} \label{SHTot} \, ,\\
		\hat H_0&=\omega_c \hat a^\dag \hat a+\omega_m \hat b^\dag \hat b \label{SH0}\, , \\ 
		\hat H_{\text{int}}&=- g_{\mathrm{lin}} \hat a^\dag \hat a (\hat b+\hat b^\dag)- g_{\mathrm{quad}} \hat a^\dag \hat a 
		(\hat b+\hat b^\dag)^2 \label{SHInt}\, , \\
		\hat H_{\text{drive}}&=\mathcal{E}(t) \hat a^\dag+\mathcal{E}^*(t) \hat a \label{SHDrive} \, .
	\end{align}

\begin{figure}[tb]
		\includegraphics[width=\columnwidth]{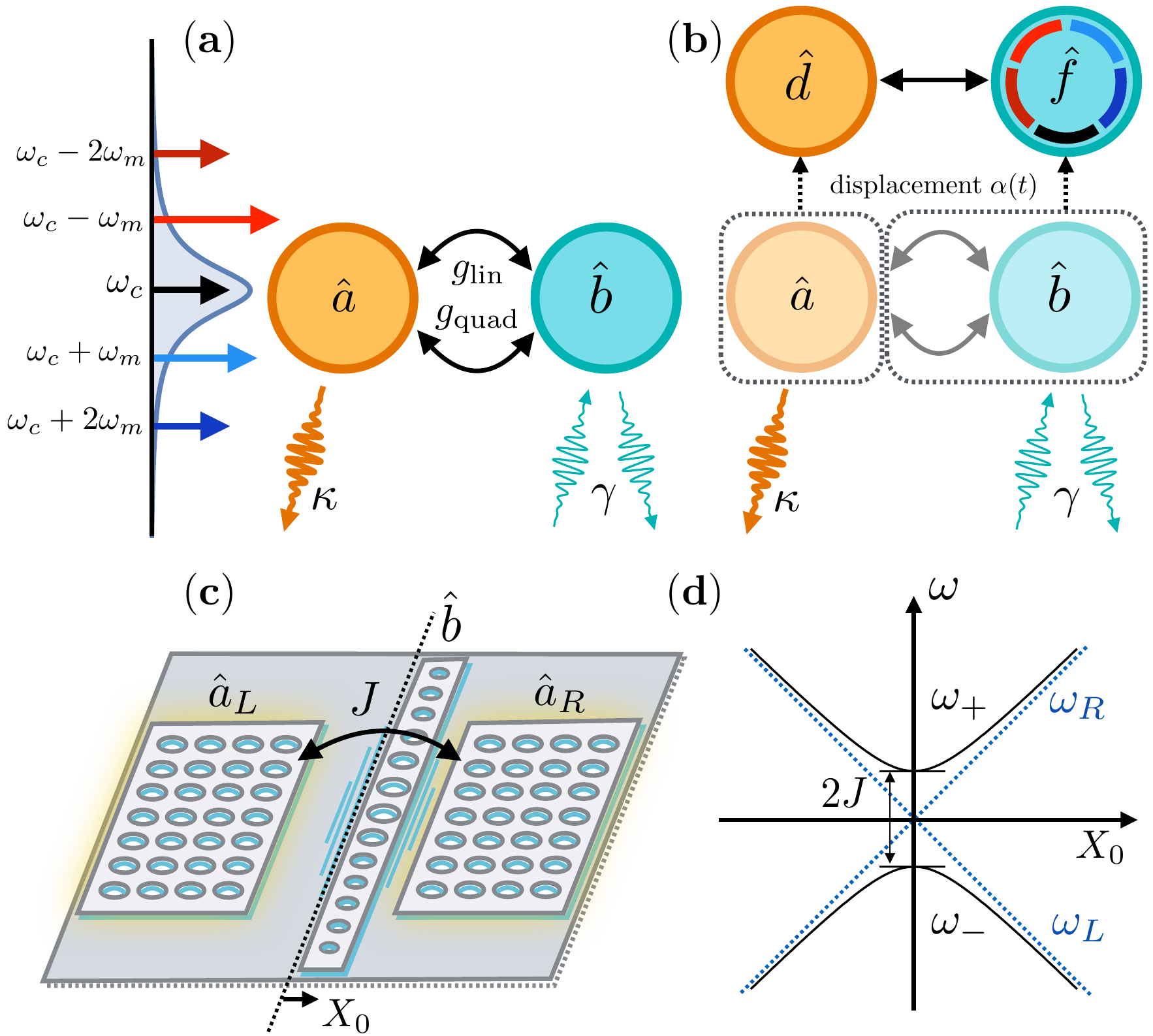}
		\caption{({\bf a}) Linear-and-quadratic optomechanical interaction between a mechanical resonator ($\hat b$) and a cavity mode ($\hat a$) 
		[Eq.~\eqref{SHTot}], with the cavity  coherently driven on multiple frequencies.  ({\bf b}) In a displaced frame and after discarding fast oscillating terms, 
		the mechanical mode $\hat b$ gets `dressed' by the linear-and-quadratic interaction (nonlinear operator $\hat f$) and couples via a beam-splitter interaction
		to the cavity fluctuation $\hat d$ [Eq.~\eqref{SRWA}], thus implementing the model of Eq.~\eqref{BS}. 
		The different contributions to $\hat f$ are determined by the relative strengths and phases among the drives, as symbolized by the inner circle. 
		({\bf c}) An optomechanical crystal implementation of a tunable linear-and-quadratic coupling. Photons of the confined photonic mode $\hat a_{L,R}$  
		hop through a mechanical resonator ($\hat b$) and hybridize into supermodes $\omega_{\pm}$ delocalized over the two slots. 
		For a central beam equidistant from the two slots ($X_0=0$), a purely quadratic 
		optomechanical coupling is realized, while with a tunable offset of the beam displacement $X_0$, arbitrary linear-and-quadratic admixtures 
		can be achieved [Eqs.~\eqref{GLin} and~\eqref{GQuad}].  ({\bf d}) Hybridization of the right and left modes 
		into supermodes and avoided crossing.   
		\label{fig:PlotOptomech}}
	\end{figure}

	The first expression, Eq.~\eqref{SH0}, collects the free oscillating terms, where $\hat a$ ($\hat b$) describes the cavity (mechanical) mode with frequency $\omega_c$ ($\omega_m$). Eq.~\eqref{SHInt} describes the linear and the quadratic optomechanical interaction with single-photon coupling strength $g_{\mathrm{lin}}$ and $g_{\mathrm{quad}}$, respectively, while the last term represents a linear drive consisting of several frequencies, i.e., $\mathcal{E}(t)=\sum_k \epsilon_k e^{-i \omega_k t}$. In the following, we will dub linear (quadratic) the term in Eq.~\eqref{SHInt} proportional to $g_{\mathrm{lin}}$ ($g_{\mathrm{quad}}$). Furthermore, the cavity is in contact with an effective zero-temperature reservoir, whereas the mechanical oscillator with a bath that determines a finite thermal occupancy $\bar n$~\cite{walls2007quantum}. Assuming for both processes the Markovian limit, the noise correlation functions  are given by $\nobreak{\langle \hat{a}_{\text{in}}(t)\hat{a}_{\text{in}}^{\dagger}(t')\rangle= \delta (t-t')\, ,}\nobreak{\langle \hat{a}_{\text{in}}^{\dagger}(t)\hat{a}_{\text{in}}(t')\rangle=0 \, ,}$ and $\nobreak{\langle \hat{b}_{\text{in}}(t)\hat{b}_{\text{in}}^{\dagger}(t')\rangle= (\bar n +1) \delta (t-t')\, },\nobreak{\langle \hat{b}_{\text{in}}^{\dagger}(t)\hat{b}_{\text{in}}(t')\rangle=\bar n \,\delta (t-t')}$ and the associated optical and the mechanical damping rate are $\kappa$ and $\gamma$; such a configuration is depicted in \reffig{fig:PlotOptomech} ({\bf a}).
	

	Due to the strong driving, we can separate the contributions to the cavity fileld into mean field and fluctuations, i.e.~$\nobreak{\hat a(t)=\alpha(t)+\hat d(t)}$. After a transient, the cavity field is expected to follow the modulation of the drive and we may use the ansatz $\alpha(t)=\sum_k \alpha_k e^{-i \omega_k t}$. The amplitude modulation of the intra-cavity field  translates into an oscillating force acting on the mechanical element, which can also be decomposed into mean field and  fluctuations. However, if we restrict ourselves to the weak coupling limit $\vert g_{\mathrm{lin}(\mathrm{quad})} \alpha_k \alpha_l\vert \ll \vert \omega_k-\omega_l\vert$, $k\neq l$, (we will see later that $\vert \omega_k-\omega_l\vert$ is of the order of the mechanical frequency $\omega_m$), to a good approximation we can set the mechanical mean field to zero and the (stationary) cavity components to $\alpha_{k,s} =\frac{-i \epsilon_k}{\kappa/2 -i \Delta_k }$, where $\Delta_k=\omega_k-\omega_c$ (see Appendix~\ref{sec:equilibrium-mean-field} for details). Under this approximation, the equations of motion can be derived from the following effective Hamiltonian (in a frame rotating with the cavity and mechanical frequencies)
	\begin{widetext}
		\begin{equation}\label{SHLin}
			\hat H=-\sum_k \left(\alpha_k  \hat d^{\dag}e^{-i\Delta_k t}+\alpha_k^* \hat d e^{i\Delta_k t}\right)\left[g_{\mathrm{lin}}\left(\hat b e^{-i\omega_m t}+\hat b^{\dag} e^{i\omega_m t}\right)+g_{\mathrm{quad}}\left(\hat b e^{-i\omega_m t}+\hat b^{\dag} e^{i\omega_m t}\right)^2 \right] \, .
		\end{equation}
	\end{widetext}
	Because of the joint presence of the linear and the quadratic coupling of the cavity mode to the same mechanical oscillator, standard linearization of the cavity field does not result in an overall bilinear Hamiltonian. We also note that linear and quadratic coupling to {\it different} cavity modes have been recently considered to obtain a tunable optomechanical nonlinearity~\cite{zhu2018controllable}. Although Eq.~\eqref{SHLin} contains the desired linear and quadratic terms, it is still time dependent. We now show how the desired interaction Eq.~\eqref{BS} can be recovered  by carefully choosing the cavity-pump detunings. We consider the following values of the detuning for the driving terms: $\Delta_1=-\omega_m\, , \Delta_2=\omega_m\, ,\Delta_3=-2\omega_m\, , \Delta_4=2\omega_m\, ,$ and $\Delta_5=0$, i.e., we  drive the cavity  on resonance and on both the first and second, blue and red, mechanical sidebands. It is easy to see that this choice makes the following processes in  Eq.~\eqref{SHLin} resonant
	\begin{equation}\label{SRWA}
		\hat H_{{\rm RWA}}= \hat d^{\dag}(G_1 \hat b +G_2 \hat b^\dagger+G_3 \hat b^2 +G_4 \hat b^{\dagger\,2} +G_5 \{\hat b,\hat b^\dagger\})  +\mathrm{H.c.}\, ,
	\end{equation}
	where  we set $G_{1(2)}=\alpha_{1(2)}g_{\mathrm{lin}}$,  $G_{3(4,5)}=\alpha_{3(4,5)}g_{\mathrm{quad}}$. This equation coincides with that of Eq.~\eqref{BS}, the quantum fluctuation $\hat d$ playing the role of the auxiliary mode, the mechanical resonator being identified with the target mode and the dressed optomechanical couplings $G_j$ with 
	the coefficients $c_j$. Informally speaking, the original mechanical mode gets dressed by the nonlinear interaction (amplified by the drives) and turns into the combination $\hat f$, as sketched in \reffig{fig:PlotOptomech} ({\bf b}). The individual terms of such combination can be independently tuned acting on the relative strength and phase among the drives, and when they match the expressions derived in Sec.~\ref{s:CubicPhase},~\ref{s:FiniteSup}, the setup cools the mechanical mode toward a nonclassical state of motion. 
	From Eqs.~\eqref{eqn:unitary-condition-2}-\eqref{eqn:unitary-condition-4} we see that the stabilization of a mechanical cubic phase state necessitates all five drives, whereas the two examples of macroscopic quantum superposition $\ket{\mathcal{C}^\Phi_n}$ and 	$\ket{\mathcal{C}^\Psi_n}$ four ($\epsilon_2=0$) and three ($\epsilon_2=\epsilon_3=0$), respectively; the approximated mechanical Fock state $\ket{\mathcal{F}^\Psi_n}$ requires three drives as well ($\epsilon_3=\epsilon_4=0$). Possible limited precision in achieving the prescribed values  as well as the detrimental effects of mechanical thermal decoherence have already been addressed in Sec.~\ref{s:Imprecision} and directly apply here. 

	In order to retrieve an interaction of the desired form, the counter-rotating terms $\hat H_{{\rm CR}}=\hat H-\hat H_{{\rm RWA}}$ have been dropped. These are given by $\hat H_{{\rm CR}}=\sum_{k=1}^4 H^{(k)}_{{\rm CR}}\EXP{i k\omega_m t}\ +\mbox{H.c.}$, with
\begin{align}
			H^{(1)}_{{\rm CR}}	&=	R(G_3 \hat d^\dagger+G_4^* \hat d)\hat b+R(G_5 \hat d^\dagger+G_5^* \hat d)\hat b^\dagger\nonumber \\
			&+R^{-1}(G_2 \hat d^\dagger+G_1^* \hat d)\hat {b^\dagger}^2+R^{-1}(G_1 \hat d^\dagger+G_2^* \hat d)\{\hat b,\hat b^\dagger\},\\
			H^{(2)}_{{\rm CR}}	&=	(G_1 \hat d^\dagger+G_2^* \hat d)\hat b^\dagger+(G_5 \hat d^\dagger+G_5^* \hat d)\hat {b^\dagger}^2\nonumber \\
			&+(G_3 \hat d^\dagger+G_4^* \hat d)\{\hat b,\hat b^\dagger\}\, ,\\
			H^{(3)}_{{\rm CR}}	&=	R(G_3 \hat d^\dagger+G_4^* \hat d)\hat b^\dagger+R^{-1}(G_1 \hat d^\dagger+G_2^* \hat d)\hat{b^\dagger}^2\, ,\\
			H^{(4)}_{{\rm CR}}	&=	(G_3 \hat d^\dagger+G_4^* \hat d)\hat {b^\dagger}^2\, ,
\end{align}
	where we introduced the ratio  $R=g_{\mathrm{quad}}/g_{\mathrm{lin}}$. From this explicit form it is apparent that a necessary condition for the rotating-wave approximation to be valid is that
	\begin{equation}\label{RWACondition}
		\left\vert  G_j \right\vert \ll \omega_m \,,\quad \left|R G_\mu\right|\ll\omega_m \quad \mathrm{and}\quad \left|R^{-1} G_\nu\right|\ll\omega_m \, ,
	\end{equation}
	for $j=1,\ldots,5$, $\mu=3,4,5$ and $\nu=1,2$.
	
	Let us summarize the requirements needed for the implementation of our scheme: the optomechanical system must couple simultaneously to the mechanical displacement and displacement squared; the quadratic coupling should be non-negligible compared to the linear one, which is a demanding requirement since the former scales as the square of the zero point motion amplitude; the validity of the rotating-wave approximation also sets a constraint on the ratio $R=g_{\mathrm{quad}}/g_{\mathrm{lin}}$. These considerations entail that we need a platform capable of a large and tunable quadratic nonlinearity. Tunable quadratic coupling can be realized in platforms as different as membrane-in-the-middle setup~\cite{thompson2007strong,sankey2010strong, purdy2010tunable}, cold atoms~\cite{karuza2012tunable}, microdisk resonators~\cite{doolin2014nonlinear} and photonic crystal cavities~\cite{paraiso2015position,kalaee2016design}. The latter, in particular, seem especially advantageous as they offer the strongest nonlinearity (together with cold atoms) and the highest tunability. In the rest of this Section we explicitly discuss the case of an optomechanical crystal
	implementation 
	and refer to Refs.~\cite{paraiso2015position} and~\cite{kalaee2016design} for a detailed discussion.

		
		The system, as schematically shown in \reffig{fig:PlotOptomech} ({\bf c}), consists of two degenerate optical modes $\hat a_L$ and $\hat a_R$ of frequency $\omega$, localized on two photonic crystal structures and coupled at a rate $J$ via photon hopping across a central mechanical  beam $\hat b$. The Hamiltonian of the three-mode optomechanical system is given by
		\begin{align}
			\hat H_{\mathrm{tot}}&=\hat H_0+\hat H_{\mathrm{hop}}+\hat H_{\mathrm{int}} \, ,\\
			\hat H_0&=\omega (\hat a_L^\dagger\hat a_L+\hat a_R^\dagger\hat a_R) +\omega_m\hat b^\dagger \hat b\,,\\
			\hat H_{\mathrm{hop}}&=J(\hat a_L^\dagger\hat a_R+\hat a_R^\dagger\hat a_L)\,,\\
			\hat H_{\mathrm{int}}&=x_{\mathrm{zpf}}(\hat b+\hat b^\dagger)(g_L \hat a_L^\dagger\hat a_L+g_R\hat a_R^\dagger\hat a_R)\,.
		\end{align}
		Due to the tunneling, these modes hybridize into supermodes $\hat a_\pm=(\hat a_L\pm \hat a_R)/\sqrt2$. As shown in Ref.~\cite{ludwig2012enhanced}, the Hamiltonian written in the supermode basis can be diagonalized by assuming a quasi-static approximation of the mechanical motion, resulting in eigenfrequencies  $\nobreak{\omega_\pm=\omega_\pm(\hat X)}$ that are given by
		\begin{equation}\label{Omegapm}
			\omega_\pm(\hat X)=\omega+g_\pm \hat X\pm\sqrt{J^2+g_{+-}^2\hat X^2} \, ,
		\end{equation}
		where we set  $\hat X=x_{\mathrm{zpf}}(\hat b+\hat b^\dagger)$ and
		\begin{equation}
			g_+=g_-=\frac{g_L+g_R}{2}\quad \mathrm{and} \quad g_{+-}=\frac{g_L-g_R}{2}
		\end{equation}
		are referred to as linear self-mode coupling and linear cross-mode coupling, respectively~\cite{kalaee2016design}. For the geometry sketched in \reffig{fig:PlotOptomech} ({\bf c}) one has $g_L= -g_R$, so that by expanding Eq.~\eqref{Omegapm} around a position equidistant from the two slots ($X_0=0$), one is left with a purely quadratic interaction with enhanced coupling $g_{\mathrm{quad}}=\frac{g_{+-}^2}{2J} x_\mathrm{zpf}^2$ [see \reffig{fig:PlotOptomech} ({\bf d})]. The enhancement follows from the fact that  $J$ can be made arbitrarily small. On the other hand, when the central beam position is not equidistant from the two crystal cavities, the expansion of the supermode frequency around  $X_0\neq0$ will lead to both a linear and a quadratic term.
		The expressions for the self- and cross-mode coupling in this case read~\cite{kalaee2016design}
				\begin{align}
		g_\pm(X_0)\approx&\, \frac{g_L+g_R}{2}\pm\frac{g_L-g_R}{2}\frac{Z}{\sqrt{Z^2+1}}\,, \\
		g_{+-}(X_0)\approx&\, \frac{g_L-g_R}{2}\frac{1}{\sqrt{Z^2+1}}\,, 
		\end{align}
		where $Z=\frac{(g_L-g_R)X_0}{2J}$,
which entail
\begin{align}
g_{\mathrm{lin}}&=g_\pm(X_0)\,x_\mathrm{zpf}\, , \label{GLin}\\
g_{\mathrm{quad}}&=\frac{g_{+-}^2(X_0)}{2J}\frac{1}{(Z^2+1)^{3/2}}\,x_\mathrm{zpf}^2\,. \label{GQuad}
\end{align}
Eqs.~\eqref{GLin} and~\eqref{GQuad} explicitly show that in a double-slotted optomechanical crystal it is possible, within the limits
set by the system's parameters, to independently tune the linear and quadratic optomechanical couplings.   
		The separation of the slots with respect to the central beam can be fine-tuned via electrostatic actuation with sub-nanometre precision, which provides extremely refined control over the ratio $R$.
 
		Since the first appearance of tunable optomechanical devices~\cite{thompson2007strong}, both theoretical and experimental interests have revolved around the suppression of the linear contribution with the long-standing goal to implement the quantum-non-demolition (QND) measurement of the phonon number operator~\cite{miao2009standard,ludwig2012enhanced}. Indeed, with the exception of Ref.~\cite{kalaee2016design}, we are not aware of any other work exploring quantitatively the competition between linear and quadratic coupling in optomechanical devices. We hope that our proposal may bring attention on this already accessible regime of cavity optomechanics. 
		As a matter of fact, the conditions required by our protocol are less demanding than those needed for the QND measurement of the  phonon number.
		 
		 For a quantitative estimate of the previous quantities based on state-of-the-art optomechanical crystals we refer to Appendix C of Ref.~\cite{brunelli2018unconditional}.


\section{conclusions}\label{s:Conclusions}
	Dissipation, when suitably harnessed, can be used as an engineering tool to steer a target system toward a desired state that may display genuine quantum features. In the context of bosonic systems, this approach has been successfully employed for the unconditional preparation of single- and two-mode squeezed states of mechanical oscillators~\cite{kienzler2015quantum,wollman2015quantum,pirkkalainen2015squeezing,lecocq2015quantum,ockeloen2018stabilized}. However, if one is limited to linear resources, i.e., if only bilinear system-reservoir couplings can be engineered,  only Gaussian steady states can be stabilized, which is currently a major limiting factor for bosonic reservoir engineering. In this work we have overcome this limitation and shown how several non-Gaussian states can be unconditionally prepared in a system with a tunable quadratic nonlinearity. Interestingly, in doing so, we have discovered  novel families of bosonic states. The focus on the unconditional nature of the protocol---thus avoiding any initialization issues---makes our proposal a `direct extension' of reservoir-engineered squeezing to the non-Gaussian realm, and hence particularly timing.

	Which states can be unconditionally prepared with linear-and-quadratic resources? The answer to this question is strikingly simple and revealing: either generalized cubic phase states~\cite{gottesman2001encoding} or families of finite superpositions of Fock states (modulo Gaussian unitary operations), depending on whether the transformation induced by the engineered reservoir is canonical or not. 
	
	In photonic systems only weakly nonlinear approximation of a cubic phase state have been realized so far due to the demanding nonlinearity. On the other hand, our approach enables the preparation of non-Gaussian mechanical cluster states and opens the perspective of universal measurement-based quantum computation in an all-mechanical solid state platform~\cite{houhou2015generation,darren2017arbitrary,houhou2018unconditional}. In cavity optomechanics, proposals for the preparation of mechanical nonclassical states usually rely on either the single-photon strong coupling~\cite{bose1997preparation,marshall2003towards}, which is still extremely weak in current experimental platforms, or on conditional operations (e.g. photon-subtraction or pulsed schemes), which are probabilistic and suffer from low efficiency~\cite{paternostro2011engineering,khan2016engineering,romero2011large,hoff2016measurement,vanner2011selective,brawley2016nonlinear}. In contrast, the optomechanical implementation of our scheme guarantees the unconditional preparation of nonclassical states and only requires weak coupling.
	It is also interesting to contrast our method to other existing proposals for the dissipative preparation of macroscopic quantum superpositions. Unlike the stabilization of Schr\"{o}dinger cat states that relies on a purely quadratic optomechanical coupling~\cite{tan2013generation,asjad2014reservoir}, our scheme does not require initialization. Linear-and-quadratic reservoir engineering thus proves superior  and yet requires the same degree of nonlinearity. Our proposal also differs from the dissipative of Ref.~\cite{abdi2016dissipative} inasmuch as it does not require to engineer an anharmonic potential.
	
In conclusion, engineering a coupling to a damped auxiliary mode that has both a linear and a quadratic component leads to qualitatively new features of the family of states that can be stabilized. The scheme enables the dissipative preparation of a plethora of nonclassical states, among which  most paradigmatic states, ranging from Fock-like states to cat-like states. Our results may be useful for quantum computation and fundamental test of decoherence.  

\section*{Acknowledgements}
We thank A. Nunnenkamp and A. Ferraro for carefully reading the manuscript and T. K. Para\"iso for helpful discussions.
M.~B.~is supported by the European Union's Horizon 2020 research and innovation programme under grant 
agreement No 732894 (FET Proactive HOT). O.~H.~acknowledges support from the 
SFI-DfE Investigator programme (grant 15/IA/2864), the EU Horizon2020 
Collaborative Project TEQ (grant agreement No 766900) and from the EPSRC 
project EP/P00282X/1. 

\appendix


\section{Complete expressions of  finite superpositions Eqs.~(\ref{eqn:family-1}),~(\ref{eqn:family-2}),~(\ref{FirstCatLike}),~(\ref{SecondCatLike}) and (\ref{FockLike})}\label{sec:expressions}

Below we report the complete expressions of the two families of unconditionally pure states studied in Sec.~\ref{s:FiniteSup}
and the special cases of Sec.~\ref{s:ExFiniteSup}
		
			The first family of states addressed in  Sec.~\ref{s:FirstFamily} is characterized by the wave function
			\begin{equation}
				\varphi(x)\propto\ (x+x_1)^\epsilon\ \EXP{-\alpha(x-z_1)^2}\ ,
			\end{equation}
			where $x_1$, $z_1$, $\alpha$ and $\epsilon$ are given by
			\begin{eqnarray}
				x_1			&=&	\frac{1}{\sqrt2}\ \frac{c_1-c_2}{c_3-c_4}\ ,\\
				z_1			&=&	\sqrt2\ \frac{c_1c_4-c_2c_3}{c_3^2-c_4^2}\ ,\\
				\alpha		&=&	\frac12\ \frac{c_3+c_4}{c_3-c_4}\ ,\\
				\epsilon	&=&	-\frac{1}{2}-\frac{(c_1c_4-c_2c_3)(c_1-c_2)}{(c_3-c_4)^3}\ . \label{epsilon}
			\end{eqnarray}
		The normalization factor of the state in \refeq{eqn:family-1} is given by
			\begin{equation}
				\mathcal{N}_{\Phi_n}=\left(n!\sqrt{\pi}\ \laguerre_n^{(-\frac12)}(-\lambda^2)\right)^{-\frac12}\,,
			\end{equation}
where
\begin{equation}
\lambda=\sqrt{2\alpha}(x_1+z_1)\,
\end{equation}
and the condition $\epsilon \stackrel{!}{=}n\in \mathbb{N}$ has been enforced.

			The second family of Sec.~\ref{s:SecondFamily} is characterized by the wave function
			\begin{equation}
				\psi_n(x)\propto\EXP{-\alpha'(x-x_2)^2}\ \Hermite_\eta\left(\tau(x-z_2)\right)\ ,
			\end{equation}
			where $x_2$, $z_2$, $\alpha'$, $\eta$ and $\tau$ are given by
				\begin{eqnarray}
					x_2			&=&	\frac{c_1 c_4+c_2 c_3-(c_1+c_2)c_5}{2\sqrt2\left(c_5^2-c_3 c_4\right)}\, ,\\
					z_2			&=&	\frac{c_2\left(c_3-c_5+\sqrt{c_5^2-c_3 c_4}\right)}{\sqrt2\ \sqrt{c_5^2-c_3 c_4}\left(c_3-c_4+2\sqrt{c_5^2-c_3 c_4}\right)}\nonumber\\
								&+& \frac{c_1\left(c_4-c_5-\sqrt{c_5^2-c_3 c_4}\right)}{\sqrt2\ \sqrt{c_5^2-c_3 c_4}\left(c_3-c_4+2\sqrt{c_5^2-c_3 c_4}\right)}\,,\\
					\alpha'		&=&	\frac{c_3-c_4+2\sqrt{c_5^2-c_3 c_4}}{2(c_3+c_4-2c_5)}\, ,\\
					\tau			&=&	\frac{\sqrt2\ \left(c_5^2-c_3 c_4\right)^{\frac14}}{\sqrt{c_3+c_4-2c_5}}\, ,\\
					\eta		&=&	\frac{c_1^2 c_4+c_2^2 c_3-2 c_1 c_2 c_5-4(c_5^2-c_3 c_4)^{\frac32}}{8(c_5^2-c_3 c_4)^{\frac32}}\,.
				\end{eqnarray}
The normalization factor in Eq.~\eqref{eqn:family-2} is given by
\begin{widetext}
			\begin{equation}
				\mathcal{N}_{\Psi_n}=\pi^{-\frac14}\left[\sum\limits_{k=0}^n(-1)^k\binom{n}{k}^2(n-k)!(2s^2)^{n-k}\Hermite_k^2\left(\frac{is u}{\sqrt{s^2-1}}\right)\right]^{-\frac12}\ ,
			\end{equation}
\end{widetext}
where we set
\begin{equation}
s=\frac{\tau}{\sqrt{2\alpha'}}\quad\mathrm{and}\quad u=\frac{1}{\sqrt{2\alpha'}}(x_2-z_2)\,
\end{equation}
and the condition $\eta \stackrel{!}{=}n\in \mathbb{N}$ has been enforced.

For completeness, we also give the full expressions of Eq.~\eqref{FirstCatLike}, Eq.~\eqref{SecondCatLike} and Eq.~\eqref{FockLike}
studied in Sec.~\ref{s:ExFiniteSup}, which are respectively given by  

\begin{align}
			\ket{\mathcal{C}^\Phi_n}=&\frac{n!}{\sqrt{(2n-1)!!}}\sum\limits_{m=0}^{\lfloor \frac{n}{2}\rfloor}\frac{1}{2^{m} m! \sqrt{(n-2m)!}}\ \ket{n-2m}, \\
			\ket{\mathcal{C}^\Psi_n}=&\sqrt{\tfrac{n!}{_2F_1\left(\frac{1-n}{2},\frac{-n}{2};1;\frac14\right)}} \sum\limits_{m=0}^{\lfloor \frac{n}{2}\rfloor}\frac{4^{-m}}{ m! \sqrt{(n-2m)!}}\ \ket{n-2m}, \\
			\ket{\mathcal{F}_n^{\Psi}}=&\frac{1}{\sqrt{_2F_1\left(-n,-n;1;d_n^{-2}\right)}} \sum_{k=0}^n \binom{n}{k}d_n^{\,-k} \ket{k},
\end{align}
with $d_n=\frac{(1-\zeta)}{4\zeta}\sqrt{\zeta(2n+1)}$ and $\zeta=\tanh r\equiv \frac{\vert c_2\vert}{\vert c_1\vert}$.


\section{Derivation of the Wigner functions Eqs.~(\ref{WCubic}),~(\ref{WPhi}) and~(\ref{WPsi})}\label{sec:wigner-function-calculation}
	In this Appendix we  derive the expressions of the Wigner function of the states reported in the main text. 
	\subsection{Wigner function of the cubic phase state}
		We aim at finding the Wigner function associated to the momentum-squeezed cubic phase state (given in \refeq{eqn:cubic-phase-state-momentum}). Thanks to the relation $\hat \Gamma(\gamma)\sqz(-r)=\sqz(-r)\Gamma(\gamma\EXP{3r})$, this latter can be rewritten as
		\begin{equation}
			\ket{\gamma,r}=\sqz(-r)\hat \Gamma(\gamma\EXP{3r})\ket{0}\equiv\sqz(-r)\ket{\gamma}\ ,
		\end{equation}
		with $\ket{\gamma}=\hat\Gamma(\gamma)\ket{0}$. By writing the cubic phase state in this form, the Wigner function for the state \ket{\gamma,r}, denoted $W_{\gamma,r}$, can be readily obtained from that of the state \ket{\gamma}, that we denote $W_\gamma$, by taking in to account the symplectic transformation corresponding to the squeezing operation $\sqz(-r)$ as follows
		\begin{equation}\label{eqn:wigner-cubic-symplectic-transform}
			W_{\gamma,r}(x,p)=W_\gamma(\EXP{-r}x,\EXP{r}p)\ .
		\end{equation}

		The Wigner function associated to \ket{\gamma} is defined as \cite{wigner1932quantum}
		\begin{equation}\label{eqn:wigner-functon-gamma}
			W_\gamma(x,p)=\frac{1}{\pi}\int\limits_{-\infty}^\infty\diff y\ \EXP{2ipy} \phi_\gamma(x+y)^*\phi_\gamma(x-y)\ ,
		\end{equation}
		where $\phi_\gamma(x)$ is the corresponding wave function,
		\begin{equation}
			\phi_\gamma(x)\equiv\braket{x}{\gamma}=\pi^{-\frac14}\EXP{-\frac12 x^2+i\gamma x^3}\ .
		\end{equation}
		By pluging in this expression in \refeq{eqn:wigner-functon-gamma} we obtain the following formula:
		\begin{equation}
			W_\gamma(x,p)=\frac{\EXP{-x^2}}{\pi^{\frac32}}\int\limits_{-\infty}^\infty\diff y\ \EXP{2i(p-3\gamma x^2)y-y^2-2i\gamma y^3}\ ,
		\end{equation}
		and after evaluating this integral\footnote{This integral is a Fourier transform of the product of a Gaussian function and a cosine function of cubic argument, which may be calculated for instance using the convolution theorem \cite{bracewell1986fourier}. In general, we have: $\int\limits_{-\infty}^\infty\diff y\ \EXP{ixy-\frac12 y^2+i\gamma \frac{y^3}{3}}=2\pi|\gamma|^{-\frac13}\EXP{\frac{1}{12\gamma^2}}\EXP{\frac{x}{2\gamma}}\Ai\left[\gamma^{-\frac13}\left(x+\frac{1}{4\gamma}\right)\right]$
		} we obtain the Wigner function for the state \ket{\gamma},
	\begin{equation}
				W_\gamma(x,p)=\mathcal{M}_{\gamma}\EXP{-\frac{p}{3\gamma}}\Ai\left[\left(\frac{4}{3\gamma}\right)^{\frac13}\left(3\gamma x^2-p+\frac{1}{12\gamma}\right)\right]\, ,
			\end{equation}
			where $\mathcal{M}_{\gamma}=\frac{\EXP{\frac{1}{54\gamma^2}}}{\sqrt{\pi}}\left(\frac{4}{3|\gamma|}\right)^{\frac13}$

		Now, by using the relation \Refeq{eqn:wigner-cubic-symplectic-transform}, we find the Wigner function of the cubic phase state of interest as follows
		\begin{widetext}
		\begin{equation}
			W_{\gamma,r}(x,p)=\mathcal{M}_{\gamma,r}\ \EXP{-\frac{p}{3\gamma\EXP{2r}}}\Ai\left[\left(\frac{4}{3\gamma}\right)^{\frac13}\left(3\gamma x^2-p+\frac{1}{12\gamma\ \EXP{4r}}\right)\right]\ ,
		\end{equation}
		\end{widetext}
		where $\mathcal{M}_{\gamma,r}$ is a normalisation coefficient given by:
		\begin{equation}
			\mathcal{M}_{\gamma,r}=\frac{\EXP{\frac{1}{54\gamma^2\EXP{6r}}}}{\sqrt{\pi}\EXP{r}}\left(\frac{4}{3|\gamma|}\right)^{\frac13}
		\end{equation}

	\subsection{Wigner function of the first family of states}
		We want to find the Wigner function for the state $\ket{\varphi_n}=\displ\left(\frac{z_1}{\sqrt2}\right)\sqz\left(\frac12\ln 2\alpha\right)\ \ket{\Phi_n}$. As before, we calculate Wigner function of \ket{\Phi_n}, then we change phase space variables according to symplectic transformation corresponding to the Gaussian $\displ\left(\frac{z_1}{\sqrt2}\right)\sqz\left(\frac12\ln 2\alpha\right)$
		\begin{equation}
			W_{\phi_n}(x,p)=W_{\Phi_n}\left(\sqrt{2\alpha}(x-z_1),\frac{p}{\sqrt{2\alpha}}\right)\ .
		\end{equation}

		The Wigner function of the state \ket{\Phi_n} is written as 
			\begin{eqnarray}
				W_{\Phi_n}(x,p)	&=&	\frac{1}{\pi}\int\limits_{-\infty}^\infty\diff y\ \EXP{2ipy} \Phi_n(x+y)^*\Phi_n(x-y)\ \nonumber\\
								&=&	(-1)^n\ \frac{|\mathcal{N}_{\Phi_n}|^2}{\pi}\ \EXP{-x^2}\nonumber\\
								&\times&\int\limits_{-\infty}^\infty\diff y\ \EXP{2ipy-y^2}\left(y^2-(x+\lambda)^2\right)^n\ \nonumber\\
								&=&	\frac{|\mathcal{N}_{\Phi_n}|^2}{\pi}\ \EXP{-x^2-p^2}\sum\limits_{k=0}^n(-1)^k\binom{n}{k}(x+\lambda)^{2(n-k)}\nonumber\\
								&\times&\int\limits_{-\infty}^\infty\diff y\ \EXP{-(y-ip)^2}\ y^{2k}\ .
			\end{eqnarray}
		By calculating the last integral we obtain
		\begin{align}
			W_{\Phi_n}(x,p)&=\mathcal{M}_{\Phi_n}\, \EXP{-x^2-p^2}\sum\limits_{k=0}^n\ (-1)^k\frac{(x+\lambda)^{2(n-k)}}{(n-k)!}\nonumber \\
			&\times \mathrm{L}_n^{(-\frac12)}(p^2)\, ,
		\end{align}
		with the normalisation factor $\mathcal{M}_{\Phi_n}$ given as
		\begin{equation}
			\mathcal{M}_{\Phi_n}=\frac{1}{\pi\ \mathrm{L}_n^{(-\frac12)}(-\lambda^2)}\ .
		\end{equation}

	\subsection{Wigner function of the second family of states}
		Here we aim at finding the Wigner function for the state $\ket{\psi_n}=\ \displ\left(\frac{z_2}{\sqrt{2}}\right)\ \sqz\left(\frac12\ln 2\alpha'\right)\ket{\Psi_n}$. Again, the Wigner function for \ket{\psi_n} is obtained from that of \ket{\Psi_n} by performing a change of phase space variables according to symplectic transformation corresponding to the Gaussian $\displ\left(\frac{z_2}{\sqrt{2}}\right)\ \sqz\left(\frac12\ln 2\alpha'\right)$
		\begin{equation}
			W_{\psi_n}(x,p)=W_{\Psi_n}\left(\sqrt{2\alpha'}(x-z_2),\frac{p}{\sqrt{2\alpha'}}\right)\ .
		\end{equation}

		We proceed, as before, calculating the Wigner function of the state \ket{\Psi_n} 
		\begin{eqnarray}
			W_{\Psi_n}(x,p)	&=&	\frac{1}{\pi}\int\limits_{-\infty}^\infty\diff y\ \EXP{2ipy} \Psi_n(x+y)^*\Psi_n(x-y)\ ,\nonumber\\
							&=&	\frac{|\mathcal{N}_{\Psi_n}|^2}{\pi}\ \EXP{-x^2}\int\limits_{-\infty}^\infty\diff y\ \EXP{2ipy-y^2} \\
							&\times& \Hermite_n\big(s y+s(x-u)\big)\Hermite_n\big(-s y+s(x-u)\big)\ ,\nonumber
		\end{eqnarray}
		which, after calculating this latter integral, evaluates to
		\begin{widetext}
		\begin{equation}
			W_{\Psi_n}(x,p)=\mathcal{M}_{\Psi_n}\, \EXP{-x^2-p^2}\sum\limits_{k=0}^n\ \binom{n}{k}^2 k!(-2s^2)^k(1-s^2)^{n-k}\left|\Hermite_{n-k}\left(\frac{s}{\sqrt{1-s^2}}(x-u+ip)\right)\right|^2\, ,
		\end{equation}
		\end{widetext}
		with the normalisation factor $\mathcal{M}_{\Psi_n}$ given as:
		\begin{equation}
			\mathcal{M}_{\Psi_n}=\sum\limits_{k=0}^n(-1)^k\binom{n}{k}^2(n-k)!(2s^2)^{n-k}\Hermite_k^2\left(\frac{is u}{\sqrt{s^2-1}}\right)\ .
		\end{equation}

\section{Mean field steady state}\label{sec:equilibrium-mean-field}
	The full Hamiltonian of the system is
	\begin{eqnarray}
		H_{\mbox{sys}}	&=&	\omega a^\dagger a+\Omega b^\dagger b+a^\dagger a\left[g_0^{(1)}\left(b+b^\dagger\right)+g_0^{(2)}\left(b+b^\dagger\right)^2\right]\nonumber\\
						&&	+\epsilon(t) a^\dagger+\epsilon^*(t) a\ .
	\end{eqnarray}
	The equations of motion of the degrees of freedom are
	\begin{eqnarray}
		\dot{a}	&=&	-i[a,H_{\mbox{sys}}]-\frac12 k a +\sqrt{\kappa}\ a_{\mbox{in}}\nonumber\\
				&=&	-\left(\frac12\kappa+i\omega\right)a-ia\left[g_0^{(1)} (b+b^\dagger)+g_0^{(2)}(b+b^\dagger)^2\right]\nonumber\\
				&&	-i\epsilon(t)+\sqrt{\kappa}\ a_{\mbox{in}}\ ,\label{eqn:eqn-motion-a}\\[0.5cm]
		\dot{b}	&=&	-i[b,H_{\mbox{sys}}]-\frac12 k' b +\sqrt{\kappa'}\ b_{\mbox{in}}\nonumber\\
				&=&	-\left(\frac12\kappa'+i\Omega\right)b-ia^\dagger a\left[g_0^{(1)}+2g_0^{(2)}(b+b^\dagger)\right]\nonumber\\
				&&	+\sqrt{\kappa'}\ b_{\mbox{in}}\ .\label{eqn:eqn-motion-b}
	\end{eqnarray}
	Here we are interested in finding the steady state of the mean fields. We write the operators as the sum of a classical field and a quantum fluctuation,
	\begin{eqnarray}
		a	&\longrightarrow&	a+\alpha\ ,\\
		b	&\longrightarrow&	b+\beta\ ,
	\end{eqnarray}
	then we substitute these two latter relations in the equations of motion, take the average and invoke a mean field approximation. Since the quantum fluctuations and noise terms have zero mean values, the resulting equations of motion will involve only the classical fields,
	\begin{eqnarray}
		\dot{\alpha}	&=&	\left(-\frac{\kappa}{2}-i\omega\right)\alpha-i\alpha\left[g_0^{(1)}\left(\beta+\beta^*\right)+g_0^{(2)}\left(\beta+\beta^*\right)^2\right]\nonumber\\
						&&	-i\epsilon(t)\ ,\label{eqn:eq-motion-alpha}\\
		\dot{\beta}		&=&	\left(-\frac{\kappa'}{2}-i\Omega\right)\beta-i|\alpha|^2\left[g_0^{(1)}+2g_0^{(2)}(\beta+\beta^*)\right]\ .\label{eqn:eq-motion-beta}
	\end{eqnarray}
	
	Now, we consider a multi-tone driving field,
	\begin{equation}
		\epsilon(t)=\sum\limits_k \epsilon_k\EXP{-i\omega_k t}\ .
	\end{equation}
	The mean field $\alpha$ will follow the same dynamics \cite{milburn2011introduction}, and hence we write
	\begin{equation}\label{eqn:ansatz-alpha}
		\alpha(t)=\sum_k\alpha_k\EXP{-i\omega_k t}\ ,
	\end{equation}
	for some complex amplitudes $\alpha_k$. Substituting this equation in \refeq{eqn:eq-motion-alpha}, we obtain:
	\begin{widetext}
	\begin{equation}
		\sum\limits_k -i\omega_k \alpha_k\EXP{-i\omega_k t}=\sum\limits_k\EXP{-i\omega_k t}\left\{\alpha_k\left[-\frac{\kappa}{2}-i\left(\omega+g_0^{(1)}(\beta+\beta^*)+g_0^{(2)}(\beta+\beta^*)^2\right)\right]-i\epsilon_k\right\}\ .
	\end{equation}
	\end{widetext}
	By matching the same frequency components we get:
	\begin{equation}
		\alpha_k=\frac{-i \epsilon_k}{\frac{\kappa}{2} -i \left[\Delta_k-g_0^{(1)}(\beta+\beta^*)-g_0^{(2)}(\beta+\beta^*)^2 \right] }\ ,
	\end{equation}
	with $\Delta_k=\omega_k-\omega$ are the detunings of the drive with respect to the resonant frequency of mode $a$.

	The expression of $\beta$ is found in a similar manner. We substitute \refeq{eqn:ansatz-alpha} in \refeq{eqn:eq-motion-beta},
	\begin{widetext}
	\begin{equation}
		\dot{\beta}=\left(-\frac{\kappa'}{2}-i\Omega\right)\beta-i\sum\limits_{k,\ell}\alpha_k^*\alpha_\ell\EXP{i(\omega_k-\omega_\ell)t}\left[g_0^{(1)}+2g_0^{(2)}(\beta+\beta^*)\right]\ ,
	\end{equation}
	\end{widetext}
	then we neglect rotating terms by invoking the rotating wave approximation to obtain
	\begin{equation}
		\dot{\beta}=\left(-\frac{\kappa'}{2}-i\Omega\right)\beta-i\sum\limits_k\left|\alpha_k\right|^2\left[g_0^{(1)}+2g_0^{(2)}(\beta+\beta^*)\right]\ .
	\end{equation}
	The validity of the rotating wave approximation is ensured when
	\begin{equation}
		\left|g_0^{(j)}\alpha_k\alpha_\ell\right|\ll|\omega_k-\omega_\ell|\ ,
	\end{equation}
	for $j=1,2$ and $k\ne\ell$.

	In the long time limit, the mean field steady state for $\beta$ satisfies $\dot{\beta}=0$ which leads to the following expression
	\begin{equation}
		\beta=\frac{-g_0^{(1)}  \sum_k \left\vert\alpha_k \right\vert^2\left( \Omega+i\frac{\kappa'}{2}\right)}{\left(\frac{\kappa'}{2}\right)^2	+\Omega\left(\Omega+4g_0^{(2)}\sum_k \left\vert\alpha_k \right\vert^2 \right)}\ .
	\end{equation}

\bibliography{references}

\end{document}